\documentclass[preprint]{aastex}
\author{A.~Tsvetkova\altaffilmark{1}, D.~Frederiks\altaffilmark{1}, S.~Golenetskii\altaffilmark{1}, A.~Lysenko\altaffilmark{1}, P.~Oleynik\altaffilmark{1}, V.~Pal'shin\altaffilmark{2}, D.~Svinkin\altaffilmark{1}, M.~Ulanov\altaffilmark{1}, T.~Cline\altaffilmark{3,5}, K.~Hurley\altaffilmark{4}, R.~Aptekar\altaffilmark{1}}
\altaffiltext{1}{Ioffe Institute, Politekhnicheskaya 26, St.~Petersburg 194021, Russia; tsvetkova@mail.ioffe.ru}
\altaffiltext{2}{Vedeneeva 2-31, St. Petersburg, Russia}
\altaffiltext{3}{NASA Goddard Space Flight Center, Greenbelt, MD 20771, USA}
\altaffiltext{4}{Space Sciences Laboratory, University of California, 7 Gauss Way, Berkeley, CA 94720-7450, USA}
\altaffiltext{5}{Emeritus}
\usepackage{cmap}
\usepackage[english]{babel}
\usepackage{indentfirst}
\usepackage{amssymb}
\usepackage{pdflscape,natbib,color}
\usepackage{graphicx}
\usepackage{gensymb}
\usepackage{amsmath}
\slugcomment{\small{Submitted to ApJ 2017-09-17, revised 2017-10-11, accepted 2017-10-17}}
\title{THE KONUS-\textit{WIND} CATALOG OF GAMMA-RAY BURSTS WITH KNOWN REDSHIFTS. I. BURSTS DETECTED IN THE TRIGGERED MODE}
\begin{document}
\sloppy
\newcounter{cit}
\newcommand{\cititem}[1]{\refstepcounter{cit}(\arabic{cit})~\label{Gen:#1}\citealt{#1}}
\begin{abstract}
In this catalog, we present the results of a systematic study of gamma-ray bursts (GRBs) with reliable redshift estimates detected in the triggered mode of the Konus-Wind (KW) experiment during the period from 1997 February to 2016 June. 
The sample consists of 150 GRBs (including twelve short/hard bursts) and represents the largest set of cosmological GRBs studied to date over a broad energy band. 
From the temporal and spectral analyses of the sample, we provide the burst durations, the spectral lags, the results of spectral fits with two model functions, the total energy fluences, and the peak energy fluxes. 
Based on the GRB redshifts, which span the range $0.1 \leq z \leq 5$, we estimate the rest-frame, isotropic-equivalent energy and peak luminosity. For 32 GRBs with reasonably-constrained jet breaks we provide the collimation-corrected values of the energetics. 
We consider the behavior of the rest-frame GRB parameters in the hardness-duration and hardness-intensity planes, and confirm the ``Amati'' and ``Yonetoku'' relations for Type II GRBs. 
The correction for the jet collimation does not improve these correlations for the KW sample. 
We discuss the influence of instrumental selection effects on the GRB parameter distributions and estimate the KW GRB detection horizon, which extends to $z \sim 16.6$, stressing the importance of GRBs as probes of the early Universe. 
Accounting for the instrumental bias, we estimate the KW GRB luminosity evolution, luminosity and isotropic-energy functions, and the evolution of the GRB formation rate, which are in general agreement with those obtained in previous studies. 
\end{abstract}
\section{INTRODUCTION}
\label{Introduction}
Although decades have passed since the discovery of gamma-ray bursts (GRBs), many aspects of this astrophysical phenomenon remain unknown.
The major breakthrough was achieved 20 years ago, when the first redshift was measured for the GRB~970508 (\citealt{Metzger1997}) and the cosmological nature of GRB sources was firmly established.

GRB redshifts are usually measured from the emission lines, the absorption features of the host galaxies imposed on the afterglow continuum, or photometrically.
However, there are other approaches to estimate redshifts, e.g. the ``pseudo-redshift'' (pseudo-z) technique based on the spectral properties of GRB prompt high energy emission (\citealt{Atteia2003}) or searching for a minimum on the intrinsic hydrogen column density vs. redshift plane (see e.g. \citealt{Ghisellini1999}).
Considering only spectroscopic and photometric redshifts there were $\sim 450$ GRBs with reliably measured redshifts by the middle of 2016.
As of 2016, the GRB redshifts fill a range from spectroscopic $z=0.0087$ (GRB~980425; \citealt{Foley2006}) to photometric $z=9.4$ (GRB~090429B; \citealt{Cucchiara2011}) or NIR spectroscopic $z=8.1$ (GRB~090423; \citealt{Salvaterra2009}); however, they are expected to occur and be detectable out to redshifts greater than $z \approx 10$ and possibly up to $z \approx 15$--$20$ \citep{Lamb2001}.

The explosion energetics is one of the key parameters for understanding the GRB progenitors and the GRB central engine physics.
Knowing a GRB redshift one can estimate the isotropic equivalent gamma-ray energy ($E_\textrm{iso}$) as a substitute for the energy released by the central engine. Huge isotropic energy releases up to $E_\textrm{iso} \lesssim 10^{55}$~erg (e.g. GRB~080916C has $E_\textrm{iso} = 8.8 \times 10^{54}$~erg at $z = 4.35$; \citealt{Abdo2009}; \citealt{Greiner2009}) were first explained for the GRB~970508 (\citealt{Waxman1998}) by taking into account jet beaming: correction for the jet collimation decreases the energy release and peak luminosity of GRBs by orders of magnitude.
The hypothesis that GRBs are non-spherical explosions implies that, when the tightly collimated relativistic fireball is decelerated by the circumburst medium (CBM) down to the Lorentz factor $\Gamma\sim1/\theta_{\mathrm{jet}}$ (where $\theta_{\mathrm{jet}}$ is the jet opening angle), an achromatic break (jet break) should appear, in the form of a sudden steepening in the GRB afterglow
light curve, at a characteristic time $t_{\mathrm{jet}}$. Knowing $t_{\mathrm{jet}}$, the jet opening angle can be estimated (\citealt{Rhoads1997}; \citealt{Sari1999}) and the collimation-corrected
GRB energy calculated. With typical collimation angles of a few degrees, the true energy release from most GRBs is $\sim10^{51}$ ergs, on par with that of a supernova \citep{Frail2001}.

The Konus-\textit{Wind} (hereafter KW, \citealt{Aptekar1995}) experiment has operated since 1994 November and plays an important role in the GRB studies thanks to its unique set of characteristics: the spacecraft orbit in interplanetary space that provides an exceptionally stable background and continuous coverage of the full sky by two omnidirectional NaI detectors, high temporal resolution and wide energy range of the detectors ($\sim$10~keV--10~MeV, nominally). KW has triggered $\sim 4350$ times on a variety of transient events, including $\sim 2700$ GRBs, up to 2016~June; thus KW has been detecting GRBs at a rate of $\approx 120$ events per~year. Being a part of the Interplanetary Network (IPN), KW is enabling GRB localizations to be constrained by triangulation (see, e.g. \citealt{Pal'shin2013} and \citealt{Hurley2013} for details).

Thanks to the wide energy range, the GRB spectral cutoff energy (parameterized as $E_\textrm{p}$, the maximum of $E F_{E}$ spectrum) can be derived directly from the KW data and the GRB energetics
can be estimated using fewer extrapolations. Coupled with well-measured redshifts, the accurate estimates of these parameters provide an excellent testing ground for widely-discussed correlations between rest-frame spectral hardness and energetics, e.g. the ``Amati" (\citealt{Amati2002}), ``Yonetoku'' (\citealt{Yonetoku2004}) or ``Ghirlanda'' (\citealt{Ghirlanda2004}) relations.
This could facilitate using GRBs as standard candles (see e.g. \citealt{Atteia1997} or \citealt{Friedman2005}) and probing cosmological parameters with GRBs (see e.g. \citealt{Cohen1997} or \citealt{Diaferio2011}).

Here, we present a complete sample of GRBs with reliably-measured redshifts which triggered KW from 1997 February to 2016 June. The sample consists of 150 bursts and represents the largest set of GRBs with known redshifts detected by a single instrument over a wide energy range. The KW bursts observed in the waiting mode will be presented in a forthcoming catalog (Tsvetkova et al., in prep.)
We start this catalog with a brief description of the KW instrument in Section~\ref{Instrumentation}. The burst sample is described in Section~\ref{Sample}.
In Section~\ref{Analysis_Results} we present the temporal and spectral analyses of the sample, and the derived observer- and rest-frame energetics.
In Section~\ref{Discussion} we discuss the derived prompt emission parameters, the KW-specific instrumental biases, and the rest-frame properties of the KW GRBs.

All the errors quoted in this catalog are at the 68\% confidence level and are of statistical nature only.
Throughout the paper, we assume the standard $\Lambda$CDM model: $H_0 = 67.3$~km~s$^{-1}$~Mpc$^{-1}$,  $\Omega_{\Lambda} = 0.685$, and $\Omega_M = 0.315$ (\citealt{Planck2013}).
We also adopt the conventional notation $Q_k = Q/10^k$. 
\section{INSTRUMENTATION}
\label{Instrumentation}
KW is a gamma-ray spectrometer designed to study temporal and spectral characteristics of gamma-ray bursts, solar flares, soft gamma repeater (SGR) bursts, and other transient phenomena over a wide energy range from 13~keV to 10~MeV, nominally (i.e., at launch; see the end of this section).
It consists of two identical omnidirectional NaI(Tl) detectors, mounted on opposite faces of the rotationally stabilized \textit{Wind} spacecraft.
One detector (S1) points toward the south ecliptic pole, thereby observing the south ecliptic hemisphere; the other (S2) observes the north ecliptic hemisphere. Each detector has an effective area of $\sim$80--160~$\rm cm^2$, depending on the photon energy and incident angle.

In interplanetary space far outside the Earth's magnetosphere, KW has the advantages over Earth-orbiting GRB monitors of continuous coverage, uninterrupted by Earth occultation, and a steady background, undistorted by passages through the Earth's trapped radiation, and subject only to occasional solar particle events.
The \textit{Wind} distance from Earth as a function of time is presented in \citet{Pal'shin2013}; it ranges up to 5.5 light-seconds.

The instrument has two operational modes: waiting and triggered.
While in the waiting mode, the count rates are recorded in three energy windows G1~(13--50~keV), G2~(50--200~keV), and G3~(200--760~keV) with 2.944~s time resolution.
When the count rate in the G2 window exceeds a $\approx 9 \sigma$ threshold above the background on one of two fixed time-scales $\Delta T_\textrm{trig}$, 1~s or 140~ms, the instrument switches into the triggered mode, for which the waiting-mode data are also available up to $T_0$+250~s.
In the triggered mode, the count rates in the three energy windows are recorded with time resolutions varying from 2~ms up to 256~ms.
These time histories, with a total duration of $\sim 230$~s, also include 0.512~s of pre-trigger history.
Spectral measurements are carried out, starting from the trigger time $T_0$, in two overlapping energy intervals, PHA1~(13--760~keV) and PHA2~(160~keV--10~MeV), with 64 spectra being recorded for each interval over a 63-channel, pseudo-logarithmic energy scale.
The first four spectra are measured with a fixed accumulation time of 64~ms in order to study short bursts.
For the subsequent 52 spectra, an adaptive system determines the accumulation times, which may vary from 0.256 to 8.192~s depending on the current count rate in the G2 window.
The last 8 spectra are obtained for 8.192~s each.
As a result the minimum duration of spectral measurements is 79.104~s, and the maximum is 491.776~s (which is $\sim$260~s longer than the time history duration).
After the triggered-mode measurements are finished KW switches into the data-readout mode for $\sim$1~hour and no measurements are available for this time interval.

For all the bursts we used a standard KW dead time (DT) correction procedure for light curves (with a DT of a few $\mu$s) and spectra (with a DT of $\sim$42~$\mu$s).
The detector response matrix (DRM), which is a function only of the burst angle relative to the instrument axis, was computed using the \textsc{GEANT4} package (\citealt{Agostinelli2003}).
The detailed description of the instrument response calculation is presented in \citet{Terekhov1998}.
The latest version of the DRM contains responses calculated for 264 photon energies between 5~keV and 30~MeV on a quasi-logarithmic scale  for incident angles from $0\degree$ to $100\degree$ with a step of $5\degree$. The energy scale is calibrated in-flight using the 1460~keV line of $^{40}$K and the 511~keV e$^+$e$^-$ annihilation line.
The gain of the detectors has slowly decreased during the long period of operation.
The instrumental control of the gain became non-functional in 1997 and the spectral range changed to 25~keV--18~MeV for the S1 detector and to 20~keV--15~MeV for the S2 detector, from the original 13~keV--10~MeV; the G1, G2, G3, PHA1, and PHA2 energy bounds shifted accordingly.

The consistency of the KW spectral parameters and those obtained in other GRB experiments
was verified by a cross-calibration with Swift-BAT and Suzaku-WAM~\citep{Sakamoto2011a},
and in joint spectral fits with Fermi-GBM~(e.g., \citealt{Lipunov2016}).
It was shown that the difference in the spectrum normalization between KW and
these instruments is $\lesssim 20$\% in joint fits.
A more detailed discussion of the KW instrumental issues can be found in \citealt{Svinkin2016}, hereafter S16.

\section{THE BURST SAMPLE}
\label{Sample}
The sample comprises 150 GRBs with reliable redshift estimates detected by KW in the triggered mode from the beginning of the afterglow era in 1997 to the middle of 2016.
The general information about these bursts is presented in Table~\ref{generaltab}.
The first three columns contain the GRB name as provided in the Gamma-ray Burst Coordinates Network circulars\footnote{http://gcn.gsfc.nasa.gov/gcn3\_archive.html}, the KW trigger time $T_0$, and the KW trigger time corrected for the burst front propagation from \textit{Wind} to the Earth center (the geocenter time).

\begin{figure}
\center
\includegraphics[width=0.8\textwidth]{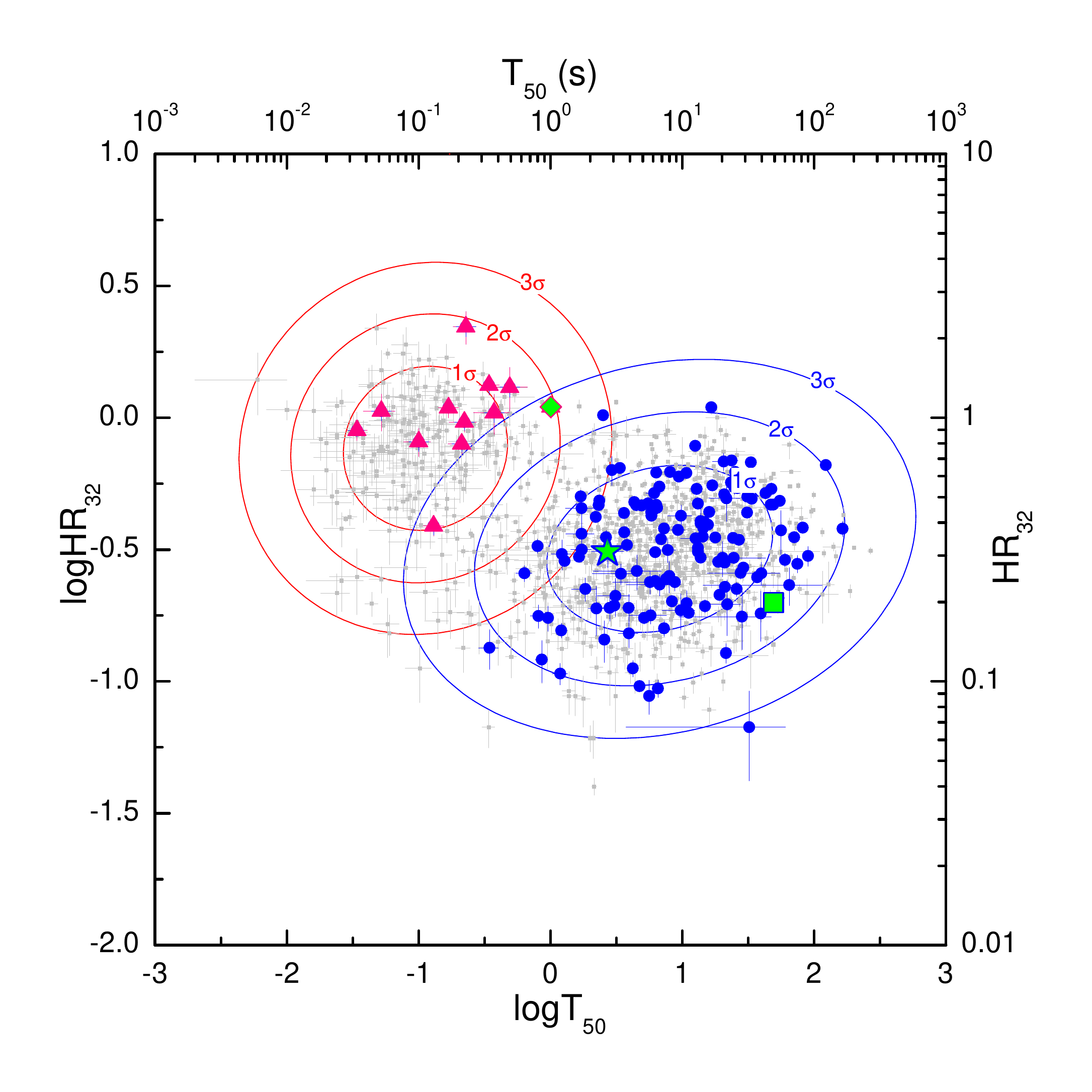}
\caption{Hardness-duration distribution of GRBs with known redshifts detected by KW in the triggered mode (Type I: red triangles; Type II: blue circles; GRB~160410A: green diamond; GRB~060614: green star, initial pulse, and green square, the whole burst). The distribution of 1143 KW bright GRBs (S16) is shown in background. This distribution is fitted by a sum of two Gaussian distributions and the contours denote 1$\sigma$, 2$\sigma$, and 3$\sigma$ confidence regions for each Gaussian distribution.}
\label{GrHRT50}
\end{figure}

The ``Type'' column specifies the burst ``physical'' classification:
Type~I, the merger-origin (\citealt{Blinnikov1984,Paczynski1986,Eichler1989,Paczynski1991,Narayan1992}), typically short/hard bursts, and Type~II, the collapsar-origin (\citealt{Woosley1993,Paczynski1998,MacFadyen1999,Woosley2006}), typically long/soft GRBs,
see, e.g., \citet{Zhang2009} for more information on this classification scheme.
According to the KW Type~I/II criteria (S16), eleven GRBs from the sample can be confidently classified as Type~I and 137 GRBs as Type~II.
Although $T_{50}\approx1.0$~s for GRB~160410A exceeds 0.6~s, a threshold used by S16 to distinguish between ``short'' and ``long'' KW GRBs,
this burst may be classified as Type~I based on its position in the hardness-duration distribution of a large sample of KW bright GRBs (Figure~\ref{GrHRT50}),
and also on its short $T_{90}\approx1.6$~s (see Section~\ref{Temporal} for definitions of $T_{50}$ and $T_{90}$).
The physical classification of GRB~060614 is unclear: a SN-less, long-duration burst (\citealt{Gehrels2006}; \citealt{Gal-Yam2006}; \citealt{Fynbo2006}; \citealt{DellaValle2006})
was suggested to be Type~I based on a low specific star-forming rate of its host galaxy (\citealt{Zhang2009});
conversely, from the KW prompt-emission analysis this GRB was classified by S16 as Type~II, that we will use in this paper.
Thus, of 150 GRBs in the sample, we designate 138 GRBs as Type~II and twelve (or 8\% of the sample) as Type~I.

The next column indicates the mission/instrument that provided the most accurate GRB localization from prompt emission observations,
thus enabling further identification of the source.
Among 150 bursts in this catalog, 103 (or $\sim$2/3) are \textit{Swift}-BAT GRBs, 13 were localized by \textit{BeppoSAX}, 14 by \textit{Fermi} (LAT and/or GBM), 8 by \textit{HETE-2},
2 by \textit{INTEGRAL}-IBIS/ISGRI, and 2 by \textit{RXTE}-ASM; for 10 GRBs, the best ``prompt'' localization was obtained with the help of triangulation by the Interplanetary Network (IPN, Hurley et al., EAS Pub Ser., 61, 459, 2013).
The ``Other obs.'' column provides the information on the burst prompt emission detections by other missions with spectrometric capabilities in hard X-ray and $\gamma$-ray domains.
The statistics of these detections are as follows: \textit{CGRO}-BATSE -- 5, \textit{HETE-2} -- 10, \textit{BeppoSAX}-GRBM -- 13, \textit{Swift}-BAT -- 102, \textit{Fermi}-GBM -- 52, and \textit{Fermi}-LAT -- 21.
The ``Det.'' and ``Inc. angle''  columns specify the KW triggered detector and the angle between the GRB direction and the detector axis (the incident angle).

The rightmost three columns of Table~\ref{generaltab} contain the redshift data.
For a number of GRBs there are several independent redshift estimates available, of which we gave a preference to spectroscopic over photometric redshift, if available;
also, results from refereed papers, which presented a detailed spectral analysis, were given higher priority over earlier GCN cirulars.
The redshift study of GRB~060121 (\citealt{deUgartePostigo2006}) revealed two probability peaks.
The main one (that we chose for this catalog, with a 63\% likelihood) places the burst at $z = 4.6 \pm 0.5$.
A secondary peak (with a 35\% likelihood) would imply that the source lies at a  $z = 1.7 \pm 0.4$.
The redshift estimate we use for GRB~150424A ($z=0.3$, \citealt{Castro-Tirado2015}) is based on the observation of a galaxy 5\arcsec (22.5 kpc at this $z$)
away from the afterglow position reported by \cite{Perley2015b}.
We note, however, that \cite{Tanvir2015} found a fainter potential host galaxy with a likely redshift of $z > 0.7$ underlying the GRB position.

\begin{figure}
\center
\includegraphics[width=0.8\textwidth]{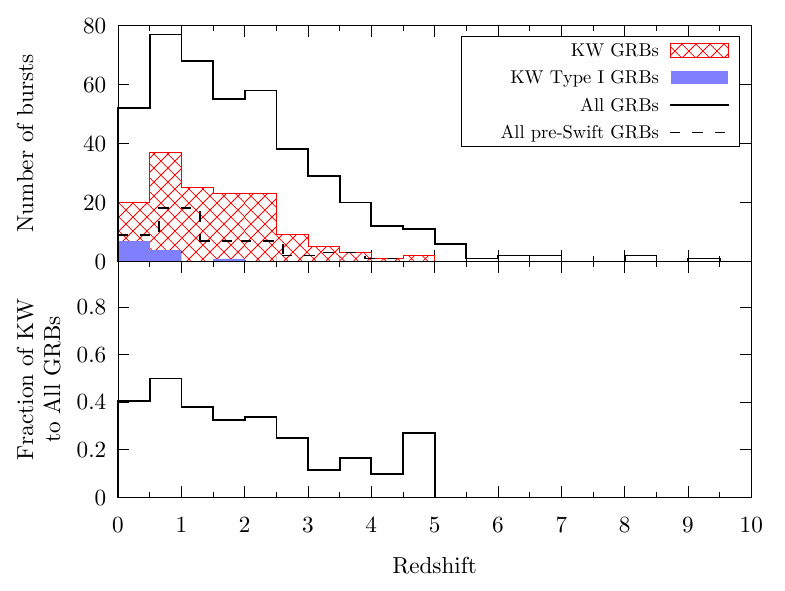}
\caption{Redshift distributions for GRBs detected up to 2016 June.}
\label{redshiftdistr}
\end{figure}

Figure~\ref{redshiftdistr} shows KW GRB redshift distributions along with those for the pre-\emph{Swift}-era GRBs and all GRB redshifts mesured to mid-2016.
The KW GRB redshifts span the range  $0.1 \leq z \leq5$ and have mean and median values of $\sim 1.5$ and $\sim 1.3$, respectively.
These statistics are comparable with those for the pre-\textit{Swift} era GRBs, whose distribution peaks at $z \sim 1$ \citep{Berger2005a},
but they are significantly lower than the \textit{Swift} era values ($\bar{z}\sim 2.3$, \citealt{Coward2013}).
The fraction of the KW-detected GRBs is $\sim$0.4--0.5 at $z<1$ and it gradually decreases with $z$, for short/hard (Type~I) bursts the fraction is $\sim$0.5. 
The absence of high-redshift bursts ($z > 5$) in the KW sample results from several instrument-specific biases discussed further in this paper.

\section{DATA ANALYSIS AND RESULTS}
\label{Analysis_Results}
\subsection{Burst Durations and Spectral Lags}
\label{Temporal}
\subsubsection{Analysis}

The total burst duration $T_{100}$, and the $T_{90}$ and $T_{50}$ durations (the time intervals which contain 5\% to 95\% and 25\% to 75\% of the total burst count fluence, respectively; see, e.g.,
\citealt{Kouveliotou1993}), were determined, in this work, using the counts in G2+G3 energy band ($\sim$80--1200~keV at present).
The soft energy band G1 was excluded from the analysis for a number of reasons, i.e.: (1) the major fraction of the GRB spectra have the peak energy of the $E F_{E}$ spectrum $E_\textrm{p} > 100$~keV and hence photons responsible for the burst energy are detected mostly in the G2 and G3 bands; (2) the KW background in G2 and G3 is very stable (in contrast to background in the soft energy range G1 ($\sim$20--80~keV) which can exhibit significant variations due to solar activity and hard X-ray transients); (3) for some bursts, an emerging X-ray afterglow may be confused with the prompt emission in G1. 

To compute the durations, a concatenation of waiting-mode and triggered-mode light curves was used.
The burst's start and end times were determined at 5$\sigma$ excess above background on time scales from 2~ms to 2.944~s in the interval from $T_0 - 200$~s to $T_0 + 240$~s (the end of the KW triggered mode record). 
In some cases, e.g., for GRB~020813, which partly overlaps in time with a solar flare, the search interval was narrowed to exclude the non-GRB event.
The background was approximated by a constant, using, typically, the interval from $T_0 - 1200$~s to $T_0 - 200$~s.

The spectral lag ($\tau_\textrm{lag}$) is a quantitative measure of spectral evolution often seen in long GRBs, when the emission in a soft detector band peaks later or has a longer decay relative to a hard band; a positive $\tau_\textrm{lag}$ corresponds to the delay of the softer emission.
To derive spectral lags we used a cross-correlation method similar to that described in ~\citet{Band1997} and \citet{Norris2000}.
The cross-correlation function (CCF) was computed between three pairs of KW energy channels: G2--G1, G3--G1, and G2--G3.
For each pair of channels (G$i$,G$j$) the peak of fourth-degree polynomial fit for the CCF was taken as $\tau_\textrm{lagGiGj}$.
The $\tau_\textrm{lag}$ error was estimated via the bootstrap approach.
To ensure the robustness of the  analysis, only bursts featuring a single emission episode, with start and end times being within the triggered mode record, were selected for the spectral lag calculations.

\subsubsection{Results}

\begin{figure}
	\center
	\includegraphics[width=0.9\textwidth]{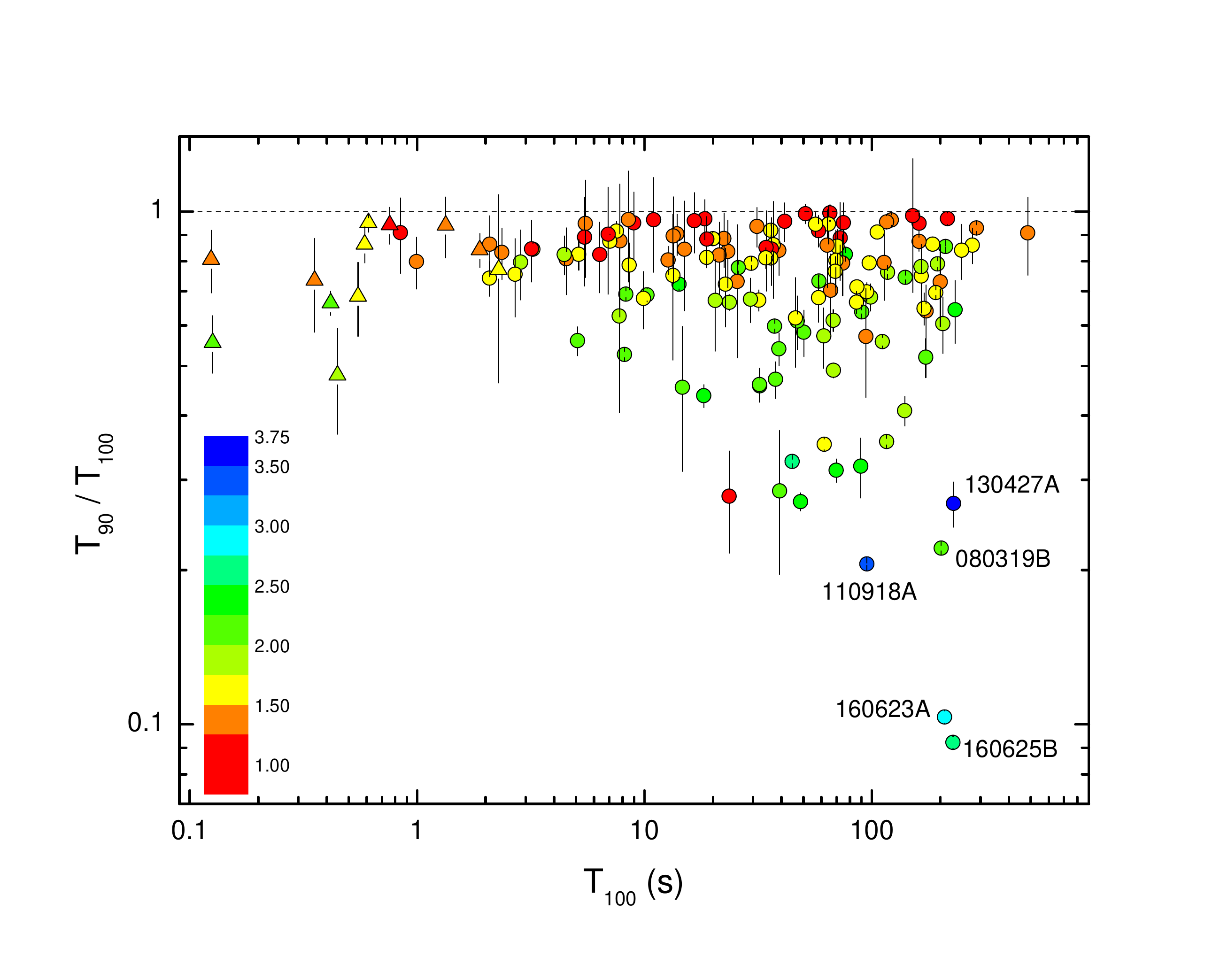}
	\caption{{\bf $T_{90}$ to $T_{100}$ ratio plotted vs. $T_{100}$. Type~I and Type~II GRBs are shown with triangles and circles, respectively. The color of each data point represents the log of the burst's trigger significance ($\sigma$).}}
	\label{T90T100}
\end{figure}

Table~\ref{temporaltab} summarizes the results of our temporal and lag analyses.
The first column contains the GRB name (see Table~\ref{generaltab}).
Next, the values of $T_{100}$, $T_{90}$ and $T_{50}$ are listed along with the corresponding start times $t_{0}$, $t_{5}$ and $t_{25}$ given relative to the trigger time $T_0$. For GRB~081203A, which was detected during the data output of GRB~081203B, no high-resolution light curves are available and, thus, only a rough estimate of $T_{100}$ is provided. 
While for weak KW GRBs $T_{100}$ and $T_{90}$ are nearly similar measures of duration (Figure~\ref{T90T100}), 
for brighter bursts $T_{100}$ becomes more sensitive to the existence of weak precursors or extended tails. 
This behavior is particularly apparent for such remarkable events as the ``naked-eye'' GRB~080319B \citep{Racusin2008}; the ultra-luminous GRB~110918A \citep{Frederiks2013}; the nearby, ultra-bright GRB~130427A \citep{Maselli2014}; 
and two recent highly energetic events, GRB~160623A \citep{Frederiks2016} and GRB~160625B (\citealt{Svinkin2016b,Zhang2017}). 
The latter burst features a precursor separated from the main episode by a long interval of quiescence and four former are characterized by slowly-decaying tails of hard X-ray emission that were bright enough to be detected in the KW G2 band for hundreds of seconds.

The last three columns of Table~\ref{temporaltab} present the spectral lags $\tau_\textrm{lagG2G1}$, $\tau_\textrm{lagG3G1}$, and $\tau_\textrm{lagG3G2}$.
For the 58 GRBs selected for the spectral lag analysis, the numbers of lags calculated are as follows: $\tau_\textrm{lagG2G1}$ (G2-G1) -- 55, $\tau_\textrm{lagG3G1}$ (G3-G1) -- 32, and $\tau_\textrm{lagG3G2}$ (G3-G2) -- 38. The missing lag values are not constrained; this may be due to a weak detection in one or both analyzed channels, or to a significant difference in a pulse shape between them.

\begin{figure}
\center
\includegraphics[width=0.8\textwidth]{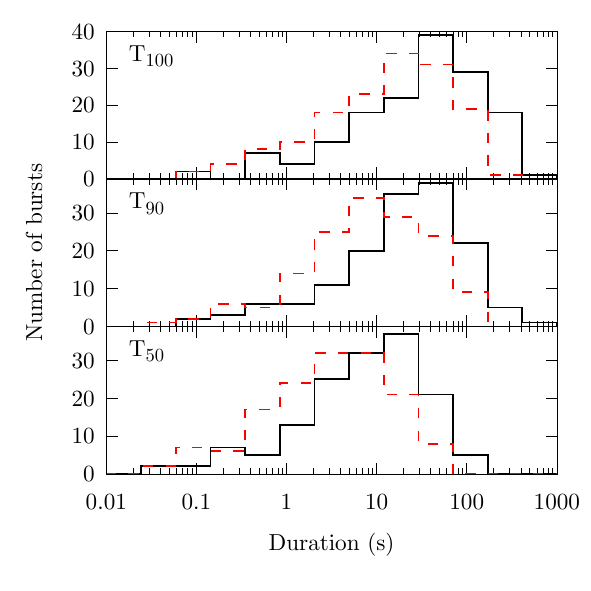}
\caption{Distributions of $T_{100}$ (top), $T_{90}$ (middle), and $T_{50}$ (bottom) in the observer- and cosmological rest frames (black solid and red dashed lines, respectively).}
\label{T100distr}
\end{figure}

Table~\ref{stattab} provides descriptive statistics of the durations and spectral lags both in the observer frame and in the cosmological rest frame.
The latter quantities are the corresponding observer-frame values scaled by the time-dilation factor 1/($1 + z$).
Figure~\ref{T100distr} presents the $T_{50}$, $T_{90}$, and $T_{100}$ observer- and rest-frame distributions.
We note that the observer-frame energy band G2+G3, in which the durations are calculated, corresponds to multiple energy bands 
in the source-frame thus introducing a variable energy-dependant factor which must be accounted for when analyzing the rest-frame durations.
The same considerations apply to the spectral lags. 

\subsection{Energy Spectra}
\label{Spectral}
\subsubsection{Analysis}

For each burst from our sample two time intervals were selected for spectral analysis:
time-averaged fits were performed over the interval closest to $T_{100}$ (hereafter the TI spectrum);
the peak spectrum corresponds to the time when the peak count rate (PCR) is reached. 
The peak spectrum accumulation time may vary from burst to burst depending on the GRB intensity and the presence of significant spectral evolution.
For 38 bursts with poor count statistics the TI and the peak spectra are measured over the same interval.

More than a dozen bursts from the sample show two or more emission episodes separated by periods of quiescence. 
In the majority of cases, all emission episodes were included to the TI spectrum. 
KW triggered on weak precursors of GRB 120716A and GRB 160625B. To maintain a reasonable signal-to-noise ratio, only the main episodes of these bursts contributed to the spectral analysis presented in this paper.

The spectral analysis was performed using XSPEC version 12.9.0 (\citealt{Arnaud1996}).
The raw count rate spectra were rebinned to have a minimum of 20 counts per channel to ensure Gaussian-distributed count statistics
and fitted using the $\chi^2$ statistic. Each spectrum was fitted by two spectral models.
The first model is the Band function (hereafter BAND; \citealt{Band1993}):
\begin{equation}
\label{eq:Band}
f(E) \propto  \left\{ \begin{array}{ll}
E^{\alpha} \exp \left(-\frac{E (2+\alpha)}{E_\textrm{p}}\right), & \quad E<(\alpha-\beta)\frac{E_\textrm{p}}{2+\alpha} \\ E^{\beta} \left[(\alpha-\beta)\frac{E_\textrm{p}}{(2+\alpha)}\right]^{(\alpha-\beta)} \exp(\beta-\alpha), & \quad E
\ge(\alpha-\beta)\frac{E_\textrm{p}}{2+\alpha}, \\ \end{array} \right.
\end{equation}
where $\alpha$ is the low-energy photon index and $\beta$ is the high-energy photon index.
The second spectral model is an exponentially cutoff power-law (CPL), parameterized as $E_\textrm{p}$:
\begin{equation}
\label{eq:CPL}
f(E) \propto E^{\alpha} \exp \left(-\frac{E (2+\alpha)}{E_\textrm{p}}\right).
\end{equation}
In the only case where both ``curved'' models result in ill-constrained fits (GRB~080413B), a simple power-law  (PL) function was used:
$f(E) \propto E^{\alpha}$.
All the spectral models were normalized to the energy flux ($F$) in the 10~keV--10~MeV range (observer frame).
The fits were performed in the energy range from $\sim 20$~keV to the upper limit of 0.5--15~MeV, depending on the presence of statistically significant GRB emission in the MeV band
and, also, on the stability of the background in the upper spectral channels which are affected, for some GRBs, by solar particles.
The parameter errors were estimated using the \textsc{XSPEC} command \textsc{error} based on the change in fit statistic ($\Delta \chi^2 = 1$) which corresponds to the 68\% confidence level (CL).

For each spectrum, we present the results for the models whose parameters are constrained (hereafter, GOOD models).
The best-fit spectral model (the BEST model) was chosen based on the difference in $\chi^2$ between the CPL and the BAND fits.
The criterion for accepting a model with a single additional parameter is a change in $\chi^2$ of at least 6 ($\Delta \chi^2 \equiv \chi_\textrm{CPL}^2 - \chi_\textrm{BAND}^2>6$).
This criterion is widely accepted for choosing between nested spectral models in GRB studies (see, e.g., \citealt{SwiftBAT1}, \citealt{Krimm2009}, \citealt{Goldstein2012}) and corresponds to an F-test chance improvement probability of $\sim 0.015$ for a reasonably good quality of fit (the reduced chi-squared, i.e. the chi-squared per d.o.f., $\chi^2_\textrm{r} \sim 1$).
It should be noted that in the KW GCN circulars a different approach is used for the best-fit model selection: BAND is preferred over CPL in the case of the constrained fit, and not dependent on $\Delta \chi^2$.

\subsubsection{Results}

The ten columns in Table~\ref{spectraltab} contain the following information: (1) the GRB name (see Table~\ref{generaltab}); (2) the spectrum type, where ``i'' indicates that the spectrum is TI, ``p'' means that the spectrum is peak; (3) and (4) contain the spectrum start time $t_\textrm{start}$ (relative to $T_0$) and its accumulation time $\Delta T$; (5) GOOD models for each spectrum ($\dag$ indicates the BEST model); (6)--(8) $\alpha$, $\beta$, and $E_\textrm{p}$; (9) $F$ (normalization); (10) $\chi ^{2}/\textrm{d.o.f.}$ along with the null hypothesis probability given in brackets.
In cases where the lower limit for $\beta$ is not constrained, the value of ($\beta_\textrm{min} - \beta$) is provided instead, where $\beta_\textrm{min} = -10$ is the lower limit for the fits.
For the best-fit values of $\beta < -4$ only the upper limits on $\beta$ are given.

Although KW high-resolution spectra do not cover the pre-trigger emission, for $\sim 2/3$ of GRBs in our sample the TI spectra include $\geqslant$90\% of the burst counts (and only for 6 bursts this fraction is $< 50\%$). A major fraction of the GRB~090812 counts, about a half of the short GRB~100206A counts, and a significant fraction of the short GRB~070714B counts were accumulated before the trigger. For these bursts, we performed the spectral analysis using both multichannel spectra and the 3-channel spectra constructed from light-curve data. Together, these spectra cover the burst $T_{100}$ interval.

\begin{figure}
\center
\tiny
\includegraphics[width=0.33\textwidth]{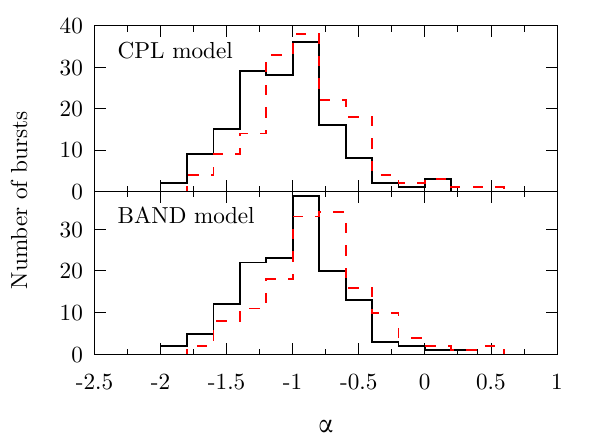}
\includegraphics[width=0.33\textwidth]{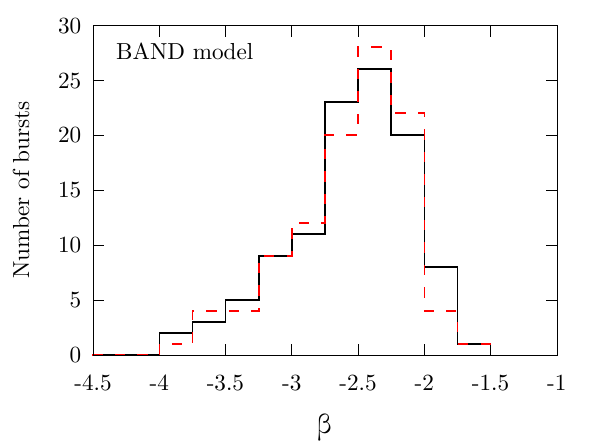}
\includegraphics[width=0.33\textwidth]{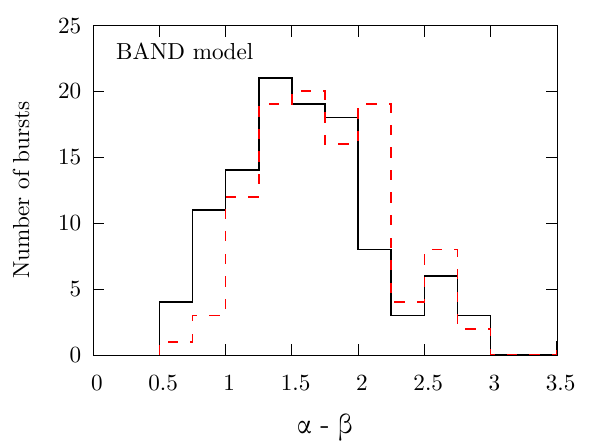}
\makebox[0.33\textwidth]{(a)}
\makebox[0.33\textwidth]{(b)}
\makebox[0.33\textwidth]{(c)}
\includegraphics[width=0.49\textwidth]{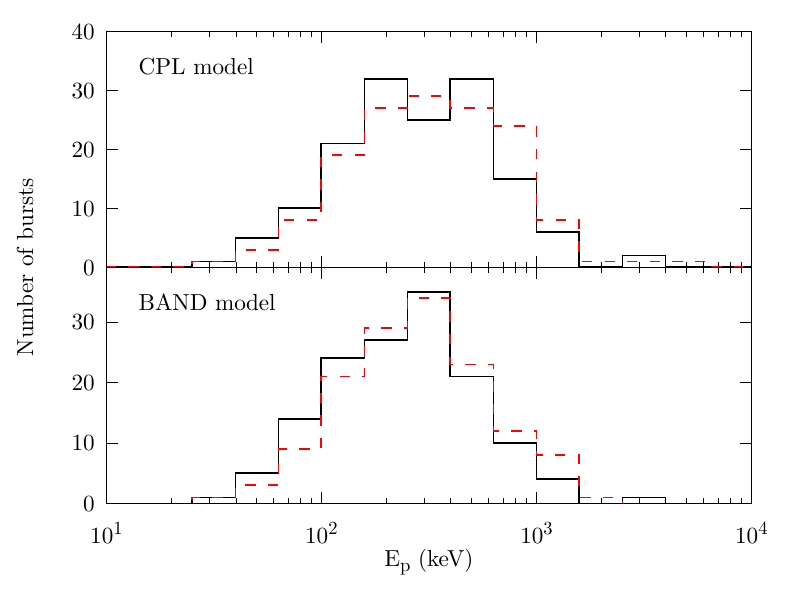}
\includegraphics[width=0.49\textwidth]{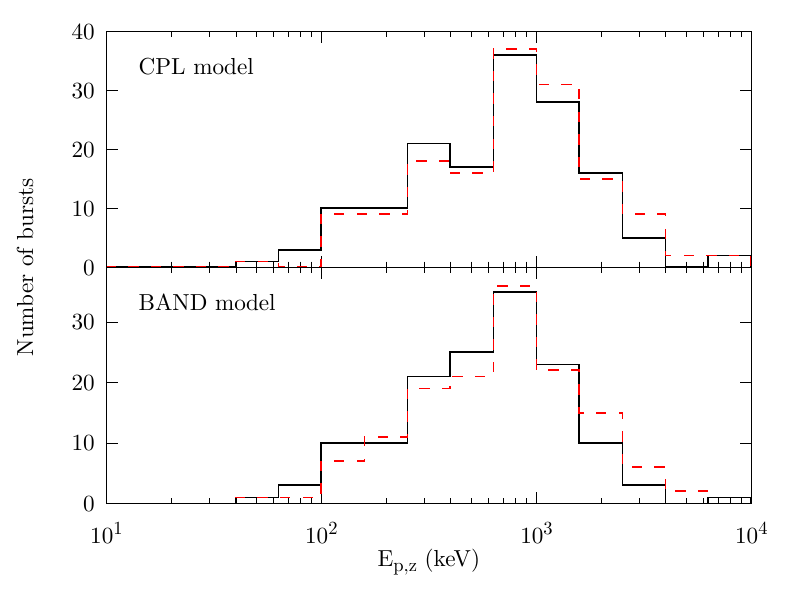}
\makebox[0.49\textwidth]{(d)}
\makebox[0.49\textwidth]{(e)}
\caption{Distributions of spectral parameters $\alpha$, $\beta$, $E_p$, and $E_\textrm{p,z}=(1+z)E_p$ for GOOD models. All the panels display the comparison between the TI spectral parameters (solid black lines) and the peak spectral parameters (dashed red lines). The panels (a),(d), and (e) also display the comparison between CPL and BAND spectral parameters.}
\label{graphspectraldistr}
\end{figure}

Figure~\ref{graphspectraldistr} shows the distributions of spectral parameters and Table~\ref{stattab} summarizes their descriptive statistics.
The CPL model's $\alpha$ for both TI and peak spectra are distributed around $\alpha \approx -1$. For the BAND model, $\alpha$ for the TI and peak spectra are distributed around $\approx -1$ and $\approx -0.85$, respectively. The high-energy photon indices $\beta$ for the TI and peak spectra are distributed around $\approx -2.5$ and $\approx -2.35$, respectively.
We found BAND to be the BEST model for 54 TI and 51 peak spectra. The remaining spectra (with the exception of GRB~080413B) were best fitted by CPL.
$E_\textrm{p}$ for the BEST model varies from $\approx 40$~keV to $\approx 3.5$~MeV (GRB~090510).
The TI spectrum $E_\textrm{p}$ ($E_\textrm{p,i}$) distributions for both spectral models peak around 250~keV, while the peak spectrum $E_\textrm{p}$ ($E_\textrm{p,p}$) distributions peak around 300~keV.
The corresponding rest-frame peak energies, $E_\textrm{p,i,z} = (1+z) E_\textrm{p,i}$ and $E_\textrm{p,p,z} = (1+z) E_\textrm{p,p}$, vary from $\approx 50$~keV  to $\approx 6.7$~MeV (GRB~090510).

\subsection{Burst Energetics}
\label{Energetics}
\subsubsection{Fluences and Peak Fluxes}
The energy fluences ($S$) and the peak energy fluxes ($F_\textrm{peak}$) were derived using the 10~keV--10~MeV energy fluxes of the BEST models for TI and peak spectra, respectively (Section~\ref{Spectral}).
Since the TI spectrum accumulation interval typically differs from the $T_{100}$ interval, a correction which accounts for the emission outside the TI spectrum was introduced when calculating $S$.
Three time scales $\Delta T_\textrm{peak}$ were used when calculating $F_\textrm{peak}$: together with two commonly utilized ones (1024~ms and 64~ms), we introduce the ``rest-frame 64~ms'' scale ($(1 + z) \cdot 64$~ms); the latter were used to estimate the rest-frame peak luminosity $L_\textrm{iso}$.
To obtain $F_\textrm{peak}$, the model energy flux of the peak spectrum was multiplied by the ratio of the PCR on the $\Delta T_\textrm{peak}$ scale to the average count rate in the spectral accumulation interval. Typically, the corrections were made using counts in the G2+G3 light curve; the G1+G2, G2 only, and G1+G2+G3 combinations were also considered depending on the emission hardness and intensity.

\subsubsection{K-correction and rest-frame energetics}
The cosmological rest-frame energetics, the isotropic-equivalent energy release $E_\textrm{iso}$ and the isotropic-equivalent peak luminosity $L_\textrm{iso}$, can be calculated, with the proper $k$-correction, as $E_\textrm{iso} = \frac{4 \pi D_\textrm{L}^2}{1+z} \times S \times k$ and $L_\textrm{iso} = 4 \pi D_\textrm{L}^2 \times F_\textrm{peak} \times k$; where $D_\textrm{L}$ is the luminosity distance. The $k$-correction to the rest-frame (see, e.g., \citealt{Bloom2001a} or \citealt{Kovacs2011} for details) is formulated in terms of spectral model energy flux $F$ as
$$k = \frac{F[E_1/(1+z), E_2/(1+z)]}{F[e_1,e_2]},$$
where [$e_1=10$~keV, $e_2=10$~MeV] is our flux calculation band in the observer frame, and [$E_1$, $E_2$] is the rest-frame ``bolometric'' energy band.
For $E_1$ we accept 1~keV and for $E_2$ -- $(1+z)  \cdot e_2=(1+z)  \cdot 10$~MeV. The latter value is higher than the widely used rest-frame limit of $10$~MeV,
since the upper boundary of the KW energy range is rather high ($> 10$~MeV) and choosing $E_2=10$~MeV would narrow the energy band of our observations.

\subsubsection{Collimation-corrected energetics}

Knowing $t_{\mathrm{jet}}$, one can estimate the collimation-corrected energy released in gamma-rays $E_{\gamma} = E_{\textrm{iso}} (1-\cos \theta_\textrm{jet})$ and the collimation-corrected peak luminosity $L_{\gamma} = L_{\textrm{iso}} (1-\cos \theta_\textrm{jet})$, where $\theta_\textrm{jet}$ is the jet opening angle and $(1-\cos \theta_\textrm{jet})$ is the collimation factor.

In the case of a CBM with constant number density $n$, hereafter HM, $\theta_\textrm{jet}$ is given by \citet{Sari1999}:
\begin{equation}
\label{eq:thetaHM}
\theta_\textrm{jet,HM}=\frac{1}{6}\left(\frac{t_\textrm{jet}}{1+z}\right)^{3/8}\left(\frac{n \eta_{\gamma}}{E_\textrm{iso,52}}\right)^{1/8},
\end{equation}
where $\eta_{\gamma}$ is the radiative efficiency of the prompt phase, $E_{\textrm{iso,52}}$ is the prompt emitted energy in units of $10^{52}$~ergs, and $t_\textrm{jet}$ is measured in days.
For calculations, we adopted canonical values $\eta_{\gamma} = 0.2$ and $n = 1$~cm$^{-3}$ (\citealt{Frail2001}).

In the case of a stellar-wind-like CBM with $n(r) \propto r^{-2}$, hereafter WM, the jet opening angle according to \citet{Li2003} is
\begin{equation}
\label{eq:thetaWM}
\theta_\textrm{jet,WM}=0.2016 \left(\frac{t_\textrm{jet}}{1+z}\right)^{1/4}\left(\frac{\eta_{\gamma} A_*}{E_{\textrm{iso,52}}}\right)^{1/4},
\end{equation}
where $A_* = (\dot{M}_\textrm{w}/(4\pi v_\textrm{w})$/($5 \times 10^{11}$~g~cm$^{-1}$) is the wind parameter, $\dot{M}_\textrm{w}$ is the mass-loss rate due to the wind, and $v_\textrm{w}$ is the wind velocity; $A_* \sim 1$ is typical for a Wolf-Rayet star.
Following \citet{Ghirlanda2007}, we assume $A_* = 1$ for all bursts with WM neglecting the unknown uncertainty of this parameter.

In this work, we only consider jet breaks that were detected either in optical/IR afterglow light curves or in two spectral bands simultaneously (e.g. in X-ray and in radio).
Among $\sim$60 jet breaks reported for KW GRBs in the literature, 32 meet this criterion (including two for Type~I bursts, GRB~051221A and GRB~030603B), and 23 of those GRBs have reasonable constraints on the CBM density profile (14 HM and nine WM).

\subsubsection{Results}

Table~\ref{energytab} summarizes observer-frame and non-collimated rest-frame energetics. The first two columns are the GRB name and $z$.
The next seven columns present the observer-frame energetics: $S$; peak fluxes on the three time scales ($F_\textrm{peak,1024}$ (1024 ms), $F_\textrm{peak,64}$ (64 ms),
and $F_\textrm{peak,64,r}$ ($(1 + z) \cdot 64$~ms)), together with start times of the intervals when the PCR is reached ($T_\textrm{peak,1024}$, $T_\textrm{peak,64}$, and $T_\textrm{peak,64,r}$). The next two columns contain $E_\textrm{iso}$ and the peak isotropic luminosity, $L_\textrm{iso}$, calculated from $F_\textrm{peak,64,r}$.
The provided $L_\textrm{iso}$ values may be adjusted to a different time scale $\Delta T$ (64 or 1024 ms) as: 
$$L_\textrm{iso}(\Delta T)=\frac{F_\textrm{peak}(\Delta T)}{F_\textrm{peak,64,r}} L_\textrm{iso}.$$

The rightmost columns provide two additional characteristics useful when the sample selection effects are considered: the bolometric energy flux corresponding to the GRB detection threshold, $F_\textrm{lim}$ (Section~\ref{Selection}); and $z_\textrm{max}$, the GRB detection horizon described in Section~\ref{Horizon}.

\begin{figure}
\center
\includegraphics[width=0.49\textwidth]{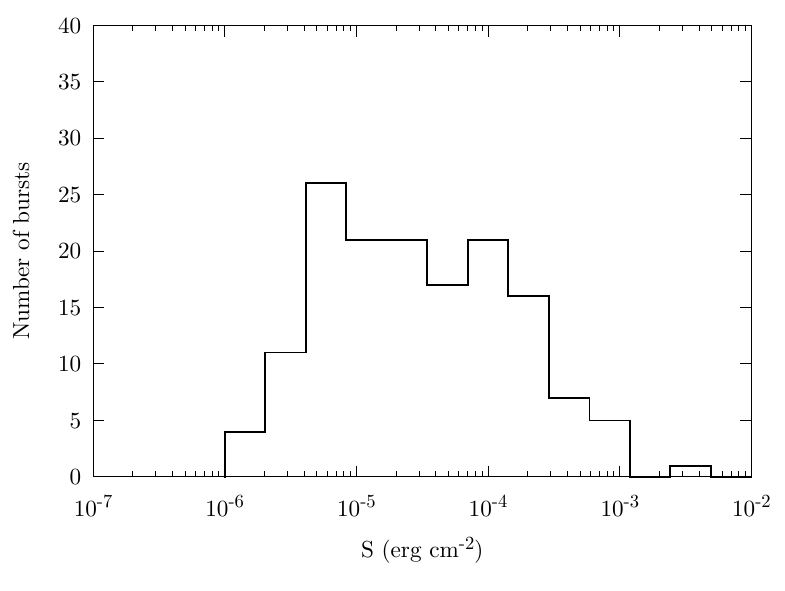}
\includegraphics[width=0.49\textwidth]{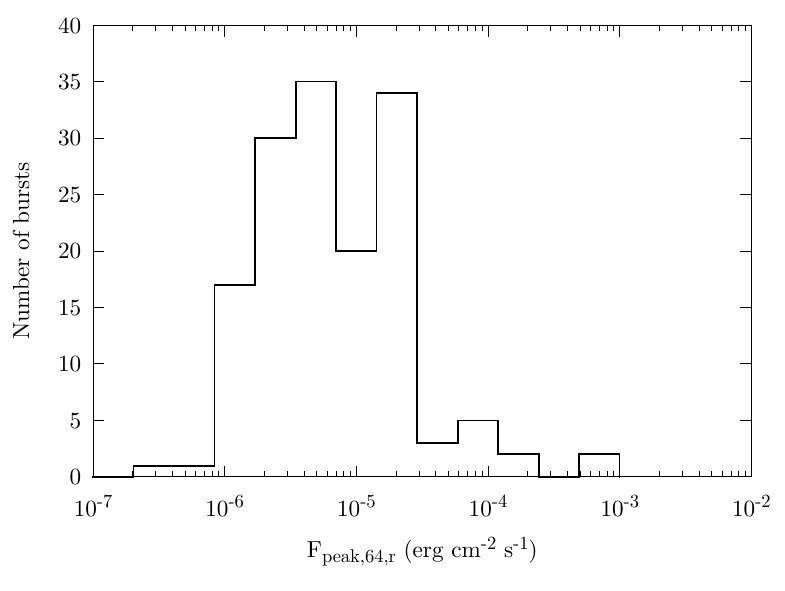}
\includegraphics[width=0.49\textwidth]{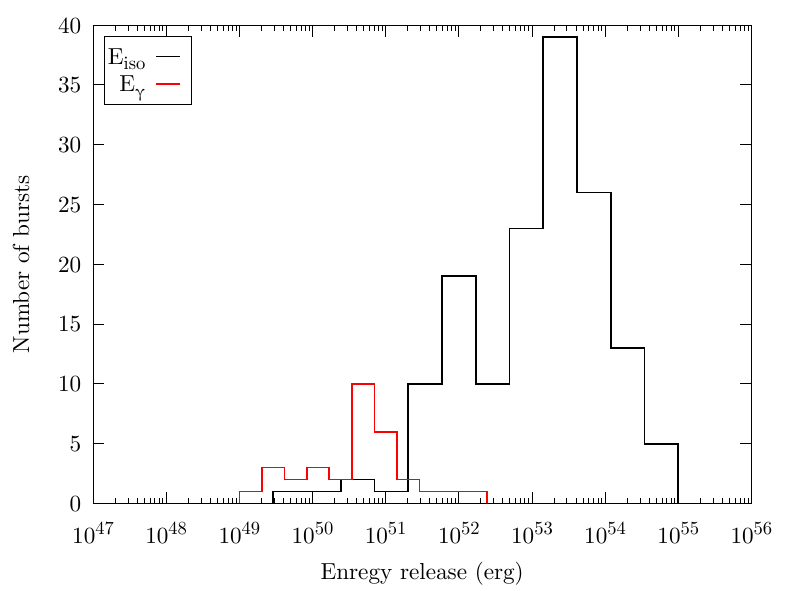}
\includegraphics[width=0.49\textwidth]{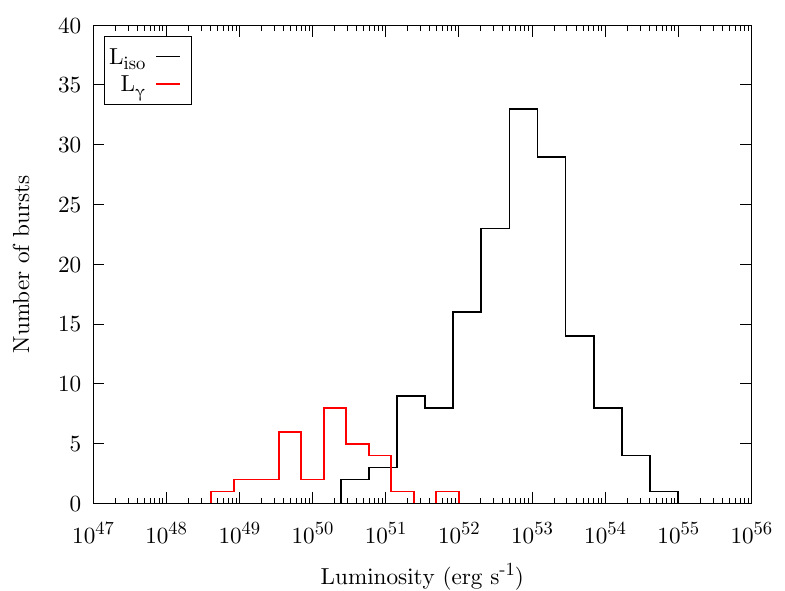}
\caption{Distributions of GRB energetics: $S$ (top left panel); $F_\textrm{peak}$ (top right panel); isotropic and collimation-corrected energy releases $E_\textrm{iso}$ and $E_\gamma$ (bottom left panel), and peak luminosities $L_\textrm{iso}$ and $L_\gamma$ (bottom right panel).}
\label{isodistr}
\end{figure}

In Figure~\ref{isodistr}, the distributions of $S$, $F_\textrm{peak,64}$, $E_\textrm{iso}$, and $L_\textrm{iso}$ are shown.
The most fluent burst in this catalog is GRB~130427A ($S = 2.86 \times 10^{-3}$~erg~cm$^{-2}$).
The brightest burst based on the peak energy flux is GRB~110918A ($F_\textrm{peak,64} = 9.02 \times 10^{-4}$~erg~cm$^{-2}$~s$^{-1}$).
The most energetic burst in terms of the isotropic energy is GRB~090323 ($E_\textrm{iso} = 5.81 \times 10^{54}$~erg).
The most luminous burst contained in this catalog is GRB~110918A ($L_\textrm{iso} = 4.65 \times 10^{54}$~erg~s$^{-1}$).

Table~\ref{collimationtab} summarizes the collimation-corrected energetics for 32 bursts with ``reliable'' jet break times.
The first column is the burst name. The next three columns specify $t_\textrm{jet}$, the CBM environment implied (HM or WM), and references to them.
The next two columns contain the derived $\theta_\textrm{jet}$ and the corresponding collimation factor,
and the last two columns present $E_\gamma$ and $L_\gamma$.
For bursts with no reasonable constraint on the CBM profile the results are given for both HM and WM.

The jet opening angles vary from $1.9\degree$ to $25.5\degree$ and the corresponding collimation factors from $5.5 \times 10^{-4}$ to $0.098$.
The brightest KW GRB in terms of both $E_{\gamma}$ and $L_{\gamma}$ is GRB~090926A ($E_{\gamma} \simeq 1.23 \times 10^{52}$~erg, $L_{\gamma} \simeq 5.50 \times 10^{51}$~erg~s$^{-1}$, $\theta_\textrm{jet} \simeq 6.20\degree$).
The distributions of $E_{\gamma}$ and $L_{\gamma}$ are shown in Figure~\ref{isodistr}.

\section{DISCUSSION}
\label{Discussion}
\subsection{This catalog \emph{vs.} previously reported KW results}
\label{Discussion_comparison_to_prev}
Preliminary results on the KW detections of bursts with known redshifts have been reported in more than a hundred GCN Circulars
and the more detailed KW GRB analyses were presented in multiple refereed publications.
Although the latter results are, with a few exceptions, statistically consistent with those reported here, the main advantage of this catalog, in comparison to the previous work,
is in the use of the unified, systematic approach to re-analyse all 150 bursts in the sample.
Particularly, GRB durations were calculated in the G2+G3 band that is less affected by the hard X-ray background variations;
this also allows one to separate the hard GRB prompt emission from the emerging X-ray afterglow.
The spectral analysis presented here gains an advantage from the most recent and accurate KW DRM; it also relies on a standard procedure for the TI spectrum interval selection based on $T_{100}$.
The burst energetics, $S$ and $F_\textrm{peak}$ are estimated, in this work, based on the BEST spectral models for TI and peak spectra, which also improves the flux calculation uncertainties.
Finally, the reported rest-frame energetics rely on the $k$-correction procedure that utilizes the full spectral band of the instrument, and they are estimated using a common set of cosmology parameters. To summarize, the results presented in this catalog form a consistent set of observer- and rest-frame GRB parameters useful for further systematic studies.

\subsection{Observer-frame spectral parameters}
\label{Discussion_spectral}
\subsubsection{The sample statistics and comparison of KW with BATSE and GBM bursts}
Although this catalog covers only a limited subset of the KW-detected GRBs ($\approx$7.5\% for the time span covered),
a discussion of the derived spectral parameter distributions may be useful for the sample characterization.

Among 138 TI spectra of long (Type~II) GRBs, 83 are best fit with the CPL model, 54 with the BAND function,
and for one GRB both ``curved'' models failed.
Similar fractions of the BEST models were obtained for the peak spectra: 86 CPL, 51 BAND, and one PL.
We found the peak spectra to be harder, in terms of $E_\textrm{p}$, as compared to the TI ones for $>$80\% of the Type~II GRBs,
consistent with the well-known GRB hardness-intensity correlation (or ``Golenetskii'' relation, \citealt{Golenetskii1983}).
Median values for the BEST model $E_\textrm{p}$ are 297~keV and 357~keV for the TI and the peak spectra, respectively.
The corresponding median $\alpha$ values are $-1.00$ and $-0.87$, and the median $\beta$ values are $-2.45$ and $-2.33$.

The case where both ``curved'' models result in ill-constrained fits is the relatively weak GRB~080413B. 
For this burst, the KW PL slope is $-2.00 \pm0.1$ ($\chi^2=49/61$~d.o.f.), suggesting a low $E_\textrm{p}$ value.
This PL slope is consistent with that derived with \textit{Swift}-BAT/\textit{Suzaku}-WAM joint fit ($-1.92 \pm0.06$, \citealt{Krimm2009}).
The best spectral model for this GRB reported by \cite{Krimm2009} is the Band function with $\alpha \simeq -1.24$, 
$\beta \simeq -2.77$, and $E_\textrm{p} \simeq 67$ (and this model is also compatible with the KW data, $\chi^2=53/62$~d.o.f.\footnote{The statistic is estimated with fixed $\alpha$, $\beta$, and $E_\textrm{p}$.}), 
that yields $E_\textrm{iso}=(2.09\pm0.28) \times 10^{52}$~erg.
Thus, the KW $E_\textrm{iso}=(3.33\pm0.61) \times 10^{52}$~erg derived for GRB~080413B from the PL fit is overestimated 
by a factor of $\sim$1.6 as compared to the more precise result of the joint BAT/WAM analysis.

Of 150 GRBs in the sample, 12 (or 8\%) are classified as short/hard (Type~I) bursts.
This fraction is half that for all KW GRBs (S16), thus reflecting the complexity of their optical identifications and redshift measurements.
All spectra of the Type~I GRBs from this catalog are fitted best by the CPL function, with median $\alpha = -0.53$ and median $E_\textrm{p} = 640$~keV.
These results are consistent with the BEST model and the spectral parameter statistics for 293 KW short GRBs given in S16.

We compared the BEST spectral parameter statistics for the whole sample with those for the BATSE~5B \citep{Goldstein2013} and \textit{Fermi}-GBM \citep{Gruber2014} catalogs.
We found the KW mean and median parameter values, for both spectral models and for both TI and peak spectra, consistent, within 68\% confidence intervals,
with the statistics given in these catalogs.
Meanwhile, we noticed some systematic differences between the instruments, e.g. the KW spectra are typically harder, in terms of $E_\textrm{p}$, than BATSE and GBM ones.
The same is true when comparing the low-energy spectral indices: the KW $\alpha$ are, on average, shallower than those reported for BATSE and GBM.
Finally, the typical KW $\beta$ are shallower than the BATSE ones, but they are steeper when compared to the typical GBM indices.
These systematic differences may be explained by the different spectral ranges of the instruments: the KW upper spectral limit ($\sim$10--15~MeV) is higher than that of BATSE ($\sim$2~MeV), thus allowing for high $E_\textrm{p}$ to be constrained better. 
In turn, the broad range of the GBM BGO detectors (up to $\sim$30~MeV) may result, for the BAND spectra, in better constrained $\beta$ and, simultaneously, smaller $E_\textrm{p}$, when compared to the typical KW ones. 
The KW-GBM spectral cross-calibration over a large sample of simultaneously observed GRBs is currently underway 
that will provide a more detailed analysis of the instrumental effects that could be affecting the scientific results from the GRB prompt emission data.

It also should be noted that the mean $E_\textrm{p}$ for the KW sample is beyond the \textit{Swift}-BAT energy range (15--150~keV),
thus emphasizing the importance of the KW detections of \textit{Swift} GRBs.

\subsubsection{Spectral indices}

The difference between low- and high-energy photon indices, $(\alpha - \beta)$, may be helpful when investigating GRB emission processes in the framework of the synchrotron shock model (SSM) through comparing the observational and theoretical values of  $(\alpha - \beta)$ to constrain the synchrotron cooling regime and infer the electron power-law index \citep{Preece2002}.
The $(\alpha - \beta)$ distribution for TI and peak spectra fitted with the BAND model is shown in Figure~\ref{graphspectraldistr} (panel c).
The fact that no obvious peak in the distributions is seen may imply a diversity of electron power-law indices and/or different SSM cases at the burst sources.
The median values of $(\alpha - \beta)$ are 1.5 and 1.6 for the TI and the peak spectra, respectively.
The peak spectrum distribution is slightly shifted towards the higher values in comparison with the TI spectrum one.

Additionally, we estimated the fraction of the bursts which violate the $-2/3$ synchrotron ``line-of-death'' (see \citealt{Preece1998} for details) and the $-3/2$ synchrotron cooling limit.
We found that the 68\% confidence intervals (CIs) for the BEST model alpha lie completely above the $-2/3$ synchrotron ``line-of-death'' for about 8\% of the TI and 21\% of the peak spectra; also, the 68\% CIs lie completely below the $-3/2$ synchrotron cooling limit for the 5\% of the TI and 2\% of the peak spectra.

\subsection{Selection effects}
\label{Selection}
Selection effects are distortions or biases that usually occur when the observational sample is not representative of the "true", underlying population.
They play a crucial role for GRBs \citep{Turpin2016, Dainotti2017}, which are particularly affected by the Malmquist bias effect that favors the brightest objects against faint ones at large distances, and these biases have to be taken into account when using GRBs as distance estimators, cosmological probes, and model discriminators.

For the sample of the KW triggered-mode GRBs with known redshifts, the selection effects fall into two categories:
the KW-specific effects, caused by its trigger sensitivity to the burst prompt emission parameters;
and the ``external'' biases arising in the process of localization and securing GRB redshifts, which are outside of the scope of this paper.

The KW triggered mode is activated when the count rate in the G2 window exceeds a $\approx 9 \sigma$ threshold above the background on one of two fixed time-scales $\Delta T_\textrm{trig}$, 1~s (applicable, with a few exceptions, to Type~II bursts in our sample) or 140~ms (the Type~I bursts).
Thus, the burst's detection significance may be characterized by a PCR to background statistical uncertainty ratio (over the corresponding $\Delta T_\textrm{trig}$).
Although the KW trigger criterion cannot be easily translated into the GRB prompt emission characteristics (e.g. duration, rise-time, spectral shape, or energy flux),
an investigation into how their combination may affect the trigger sensitivity to a specific burst may be done indirectly.

\begin{figure}
\center
\includegraphics[width=0.8\textwidth]{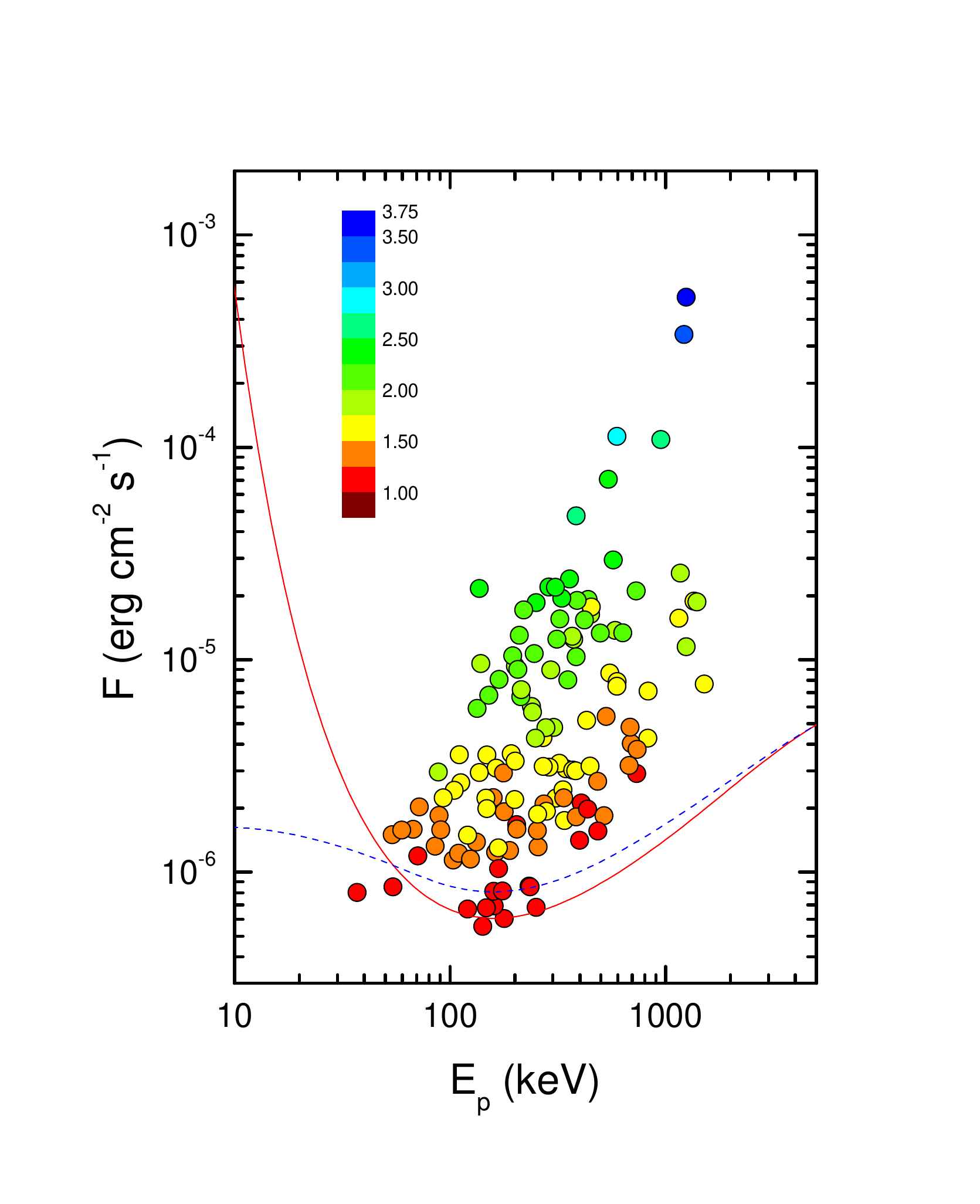}
\caption{Dependence of the limiting KW energy flux (10~keV--10~MeV) on $E_\textrm{p}$.
Calculated trigger sensitivities for CPL ($\alpha = -1$) and Band ($\alpha =-1$, $\beta = -2.5$) spectra are plotted with solid red and dashed blue lines, respectively.
$F_\textrm{peak,1024}$ (10~keV--10~MeV) vs. $E_\textrm{p,p}$ for Type~II bursts from the sample is shown by circles. The color of each data point represents the log of the burst's trigger significance ($\sigma$).}
\label{GraphSensitivity1}
\end{figure}

We estimated the energy flux sensitivity of the KW detectors following the methodology described in \citet{Band2003}.
Figure~\ref{GraphSensitivity1} presents the limiting energy flux (10~keV--10~MeV) as a function of $E_\textrm{p}$ for $\Delta T_\textrm{trig}$=1~s, for a burst incident angle $60 \degree$, and the S1 detector calibration as of mid-2015. 
As can be seen, the energy flux threshold under these assumptions is $\approx 1 \times 10^{-6}$~erg~cm$^{-2}$~s$^{-1}$ and there is a bias against the detection of soft-spectrum bursts with $E_\textrm{p} \lesssim 70-80$~keV, especially with CPL spectra, due to the instrumental selection. 
Meanwhile, the $F$--$E_\textrm{p}$ diagram stresses the lack of bright ($F \gtrsim 5 \times 10^{-6}$~erg~cm$^{-2}$~s$^{-1}$) and soft ($E_\textrm{p} \lesssim 100$~keV) GRBs, that should be easily detectable with KW. Since the lower boundary of this region is defined by GRBs with moderate-to-high detection significance, the instrumental biases do not affect the sample from this edge of the distribution. 
Thus, the apparent lack of soft/bright burst observations in the KW sample is likely due to an intrinsic GRB property (see Section~\ref{Correlations} for more discussion).

In a similar way, we calculated a limiting observer-frame energy flux for each GRB from the sample using its BEST-model spectral parameters, incident angle and detector calibration. In order to make the results more helpful for the rest-frame energy calculations, we applied $k$-corrections (Section~\ref{Energetics}) to these values using the burst redshift. The resulting bolometric limiting fluxes, $F_\textrm{lim}$, are given in Table~\ref{energytab}; the sample-mean value of $F_\textrm{lim}$ for the Type~II GRBs is $1.08 \times 10^{-6}$~erg~cm$^{-2}$~s$^{-1}$.
We note that $F_\textrm{lim}$ are calculated using the 1~s scale; when compared to peak fluxes determined on a different $\Delta T$ they should be adjusted as:
$$F_\textrm{lim}(\Delta T)=\frac{F_\textrm{peak}(\Delta T)}{F_\textrm{peak,1024}} F_\textrm{lim}.$$

Figure~\ref{GraphSensitivity2} shows the KW GRB distributions in the $E_\textrm{iso}$--$z$, $L_\textrm{iso}$--$z$, and $E_\textrm{p,z}$--$z$ diagrams.
The region in the $L_\textrm{iso}$--$z$ plane above the limit corresponding to $F_\textrm{lim} \sim 1 \times 10^{-6}$~erg~cm$^{-2}$~s$^{-1}$  may be considered free from the selection bias.
In the $E_\textrm{iso}$--$z$ plane, the selection-free region lies above the limit, corresponding to the bolometric fluence $S_\textrm{lim} \sim 3 \times 10^{-6}$~erg cm$^{-2}$.
As mentioned above, the KW detector sensitivity drops rapidly as $E_\textrm{p}$ approaches the lower boundary of the instrument's band ($\sim$20--25~keV as of 2015),
and this results in a lack of bursts below $E_\textrm{p,z,lim} \approx (1+z)^2 \cdot 25$~keV in the $E_\textrm{p,z}$--$z$ plane; the additional factor $(1+z)$ here is due to cosmological time dilation.

Finally, our sample does not exhibit any direct selection effects due to GRB duration.
However, some bursts with very gradual rising slopes may not trigger the instrument despite being bright enough to do it in other circumstances.
We estimate the number of such GRBs with known redshifts to be $\lesssim$5 (as of mid-2016); these bursts will be considered, along with other KW background-mode GRBs with known redshifts, in the second part of the catalog.

\subsection{KW GRB detection horizon}
\label{Horizon}
Knowing the maximum distance at which a particular GRB can be detected by the instrument (the GRB ``horizon'', $z_\textrm{max}$) 
may be useful in a number of applications, e.g. for deriving the $V/V_\textrm{max}$ statistic \citep{Schmidt1988} 
or for accounting for the instrumental bias when studying the ``true'' GRB energy distribution \citep{Atteia2017}.

A common approach to estimate the GRB horizon is to find a redshift $z_\textrm{max,L}$, at which the limiting isotropic luminosity $L_\textrm{iso,lim} = 4 \pi\ D_\textrm{L}^2 \times F_\textrm{lim}$, defined by the limiting energy flux estimated for the whole sample (the ``monolithic'' $F_\textrm{lim}$), starts to exceed the GRB $L_\textrm{iso}$. 
The KW trigger, however, is based on a simple photon-counting algorithm and not directly sensitive to the incident energy flux. 
Thus, the correctness of the described approach (hereafter the monolithic $F_\textrm{lim}$ method), which doesn't account for the burst-specific instrumental issues, such as trigger sensitivity to the GRB incident angle, its light-curve shape, and the shape of the energy spectrum, needs an additional confirmation.

\begin{figure}
\center
\includegraphics[width=0.8\textwidth]{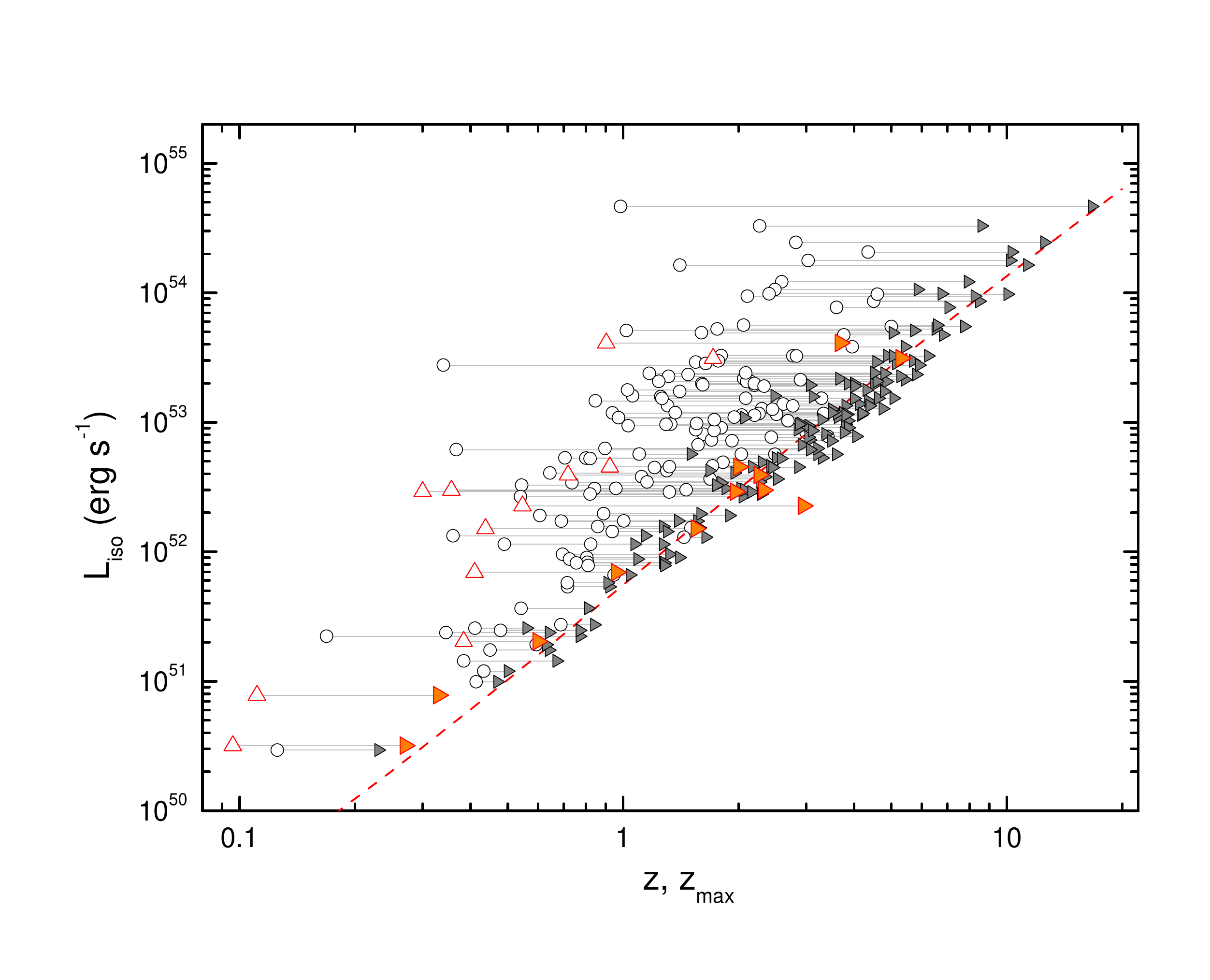}
\caption{KW GRB detection horizons plotted in the $L_\textrm{iso}-z$ plane. The solid lines connect GRBs from the sample (Type I: open triangles, Type II: open circles) and their detection horizons $z_\textrm{max}$ (filled symbols) assuming identical beaming. The limiting redshift $z_\textrm{max,L}$ defined by $F_\textrm{lim}=1\times 10^{-6}$~erg~cm$^{-2}$~s$^{-1}$ is shown by the dashed line. }
\label{Graphzmax}
\end{figure}

When evaluating how GRB detectability by KW changes when the burst source is shifted from its redshift $z$ to a more distant $z'$, at least three effects have to be accounted for.
First, the solid angle factor, which reduces (assuming identical beaming) an incident bolometric photon flux $P$ by $(D_\textrm{M}(z)/D_\textrm{M}(z'))^2$, where $D_\textrm{M}$ is the transverse comoving distance.
Second, the cosmological time dilation, which results in the light curve broadening and an additional decrease in $P$ by a factor of (1+$z'$)/(1+$z$).
Finally, the spectral cutoff, which is inherent to GRB spectra, is redshifted by the same factor, thus decreasing the fraction of $P$ within the instrument trigger band (G2).
We estimate the KW detection horizon as a redshift $z'=z_\textrm{max}$, at which the PCR in the G2 light curve drops below the trigger threshold ($9 \sigma$)
on both KW trigger scales $\Delta T_\textrm{trig}$ (140~ms and 1~s).
$\textrm{PCR}_\textrm{z'}(\Delta T_\textrm{trig})$ is calculated as
\begin{equation}
\label{eq:zmax}
\textrm{PCR}_\textrm{z'}(\Delta T_\textrm{trig}) = a \times \textrm{PCR}_\textrm{z}(a \cdot \Delta T_\textrm{trig}) \times \frac{N_\textrm{G2}(\alpha, \beta, a \cdot E_\textrm{p,p})}{N_\textrm{G2}(\alpha, \beta, E_\textrm{p,p})} \times  \left(\frac{D_\textrm{M}(z)}{D_\textrm{M}(z')}\right)^2,
\end{equation}
where $a = (1+z)/(1+z')$; $\textrm{PCR}_\textrm{z}(a \cdot \Delta T_\textrm{trig})$ is the PCR reached in the observed G2 light curve on the modified time scale;
$N_\textrm{G2}(\alpha, \beta, E_\textrm{p,p})$ is the BEST spectral model count flux in G2 calculated using the DRM; and $N_\textrm{G2}(\alpha, \beta, a \cdot E_\textrm{p,p})$ is the corresponding flux in the redshifted spectrum.
The resulting values of $z_\textrm{max}$ are given in Table~\ref{energytab} and shown in Figure~\ref{Graphzmax}.
We found that for both Type~I and Type~II GRBs, $z_\textrm{max}$ are distributed narrowly around the corresponding $z_\textrm{max,L}$ values calculated assuming the bolometric $F_\textrm{lim}=1\times 10^{-6}$~erg~cm$^{-2}$~s$^{-1}$ with the mean and $\sigma$ of the $(1+z_\textrm{max})/(1+z_\textrm{max,L})$ distribution of 1.01 and 0.12, respectively. Although in some cases $z_\textrm{max,L}$ calculated with the simple monolithic $F_\textrm{lim}$ method may differ from the more precisely evaluated $z_\textrm{max}$ by a factor of $\sim$1.5, our calculations support the general correctness of the former approach.

The most distant GRB horizon for the KW sample ($z_\textrm{max} \approx 16.6$) is reached for the ultra-luminous GRB~110918A\footnote{We found that $z_\textrm{max} \approx 7.5$ previously reported for this GRB by \citet{Frederiks2013} was miscalculated due to use of $D_\textrm{L} = (1+z) D_\textrm{M}$ instead of $D_\textrm{M}$ in Eq.~\ref{eq:zmax}. As a result, the $\textrm{PCR}_\textrm{z'}$ was underestimated by a factor of $a^2$.} at $z = 0.981$ and the second-highest ($z_\textrm{max} \approx 12.5$) is for GRB~050603 ($z=2.82$).
At $z \approx 16.6$ the age of the Universe amounts to only $\sim 230$~Myr, i.e. a burst which occurred close to the end of the cosmic Dark Ages could still trigger the KW detectors, and a thorough temporal and spectral analysis in a wide observer-frame energy range could be performed. Among the KW Type~I GRBs the highest $z_\textrm{max} \approx 5.3$ is for GRB~160410A ($z = 1.72$).

\subsection{GRB Luminosity and Isotropic-energy functions, GRB formation rate}
\label{Evolution}
Among various statistical parameters, the luminosity function as well as the cosmic formation rate of GRBs are particularly interesting.
The luminosity function (LF) is a measure of the number of bursts per unit luminosity, that provides information on the energy release and emission mechanism of GRBs.
The cosmic GRB formation rate (GRBFR) is a measure of the number of events per comoving volume and time, which can help us to understand the production of GRBs at various stages of the Universe.
While LF was originally used to study long-lasting and relatively stable astrophysical phenomena, such as stars and galaxies, the isotropic energy release function (EF,  the number of bursts per unit $E_\textrm{iso}$) can be more representative for transient phenomena, e.g., for GRBs.
The GRB EF was constructed for the first time by \citet{Wu2012} using a sample of 95 bursts with measured redshifts.
The KW sample presented in this catalog provides an excellent opportunity to test GRB LF, EF, and GRBFR on an independent dataset.

Without loss of generality, the total LF $\Phi(L_\textrm{iso},z)$\footnote{Similar reasoning may be applied to the total EF $\Psi(E_\textrm{iso},z)$} can be rewritten as $\Phi(L_\textrm{iso},z) = \rho(z)\phi(L_\textrm{iso}/g(z), \alpha_s)/g(z),$
where $\rho(z)$ is the GRB formation rate (GRBFR),
$\phi(L_\textrm{iso}/g(z))$ is the local LF, $g(z)$ is the luminosity evolution that parameterizes the correlation between $L$ and $z$,
and $\alpha_s$ is the shape of the LF, whose effect is commonly ignored as the shape of the LF does not change significantly with $z$ (e.g. \citealt{Yonetoku2004}).
Following \citet{Lloyd-Ronning2002}, \citet{Yonetoku2004}, \citet{Wu2012}, and \citet{Yu2015} we chose the functional form of $g(z) = (1 + z)^\delta$ for the luminosity evolution.
It should be noted that the isotropic luminosity evolution can be determined by either the evolution of the amount of energy per unit time emitted by the GRB progenitor or by the jet opening angle evolution (see, e.g., \citet{Lloyd-Ronning2002} for the discussion); we tested the KW sample for a correlation between the collimation factor and $z$ and found the correlation negligible (the Spearman rank-order correlation coefficient $\rho_S=-0.26$ (the corresponding $p$-value $P_{\rho_S}=0.17$) for the subsample of 30 Type~II bursts with known collimation factors).

The KW $z$--$L_\textrm{iso}$ and $z$--$E_\textrm{iso}$ samples suffer from selection effects due to the detection limit of the instrument (see Section~\ref{Selection} for details) that results in data truncation seen in Figure~\ref{GraphSensitivity2}. To estimate LF, EF, and GRBFR for the sample of 137 KW Type~II bursts we used the non-parametric Lynden-Bell $C^-$ technique (\citealt{Lynden-Bell1971}) further advanced by \citet{Efron1992} (the EP method); the details of our calculations are described in the Appendix.
The EP method is specifically designed to reconstruct the intrinsic distributions from the observed ones which account for the data truncations introduced by observational bias and includes the effects of the possible correlation between the two variables.

Applying the EP technique based on the individual (i.e. calculated for each burst independently) truncation limits
to the $z$--$L_\textrm{iso}$ plane, we found that the independence of the variables is rejected at $\tau_0\equiv\tau(\delta=0) \sim 1.7$ (where $\tau$ is a modified version of the Kendall statistic, see Appendix), and the best luminosity evolution index is $\delta_L = 1.7_{-0.9}^{+0.9}$ ($1 \sigma$ CL). Similar results were obtained using the ``monolithic'' truncation limit
$F_\textrm{lim} = 2 \times 10^{-6}$~erg~cm$^{-2}$~s$^{-1}$: $\tau_0 \sim 1.2$ and $\delta_L = 1.7_{-1.1}^{+0.9}$.

Applying the same method to the $z$--$E_\textrm{iso}$ plane and using the monolithic truncating fluence $S_\textrm{lim} = 4.3 \times 10^{-6}$~erg~cm$^{-2}$
(see Appendix for the details of $F_\textrm{lim}$ and $S_\textrm{lim}$ selection), we found that the independence of the variables is rejected at $\sim 1.6 \sigma$,
and the best isotropic energy evolution index is $\delta_E = 1.1_{-0.7}^{+1.5}$.
Thus, the estimated $E_\textrm{iso}$ and $L_\textrm{iso}$ evolutions are comparable.
The evolution PL indices $\delta_L$ and $\delta_E$ derived here are shallower than those reported in the previous studies: $\delta_L=2.60^{+0.15}_{-0.20}$ (\citealt{Yonetoku2004}),
$\delta_L=2.30^{+0.56}_{-0.51}$ (\citealt{Wu2012}), $\delta_L=2.43^{+0.41}_{-0.38}$ (\citealt{Yu2015}), and $\delta_E=1.80^{+0.36}_{-0.63}$ (\citealt{Wu2012}), albeit within errors.

After eliminating the luminosity and energy release evolution, we, following \citet{Lynden-Bell1971}, obtained the local cumulative LF and EF, $\psi (L')$ and $\psi (E')$, where $L'=L_\textrm{iso}/(1+z)^{\delta_L}$ and $E'=E_\textrm{iso}/(1+z)^{\delta_E}$.
We approximated the variance of $\psi (L')$ and $\psi (E')$ by bootstrapping the initial sample and fitted the distributions with a broken power-law (BPL) function:
$$ \psi(x) \propto \left\{ \begin{array}{ll}
x^{\alpha_1}, & \quad x \leq x_b, \\
x_b^{(\alpha_1 - \alpha_2)} x^{\alpha_2}, & \quad x > x_b, \\
\end{array} \right. $$
where $\alpha_1$ and $\alpha_2$ are the PL indices at the dim and bright distribution segments, and $x_b$ is the breakpoint of the distribution;
and with the CPL function\footnote{The CPL function definition is different here from that in Section~\ref{Spectral}}: $\psi(x) \propto x^\alpha\:\textrm{exp}(-x/x_\textrm{cut})$,
where $x_\textrm{cut}$ is the cutoff luminosity (or energy).

\begin{figure}
\center
\includegraphics[width=0.49\textwidth]{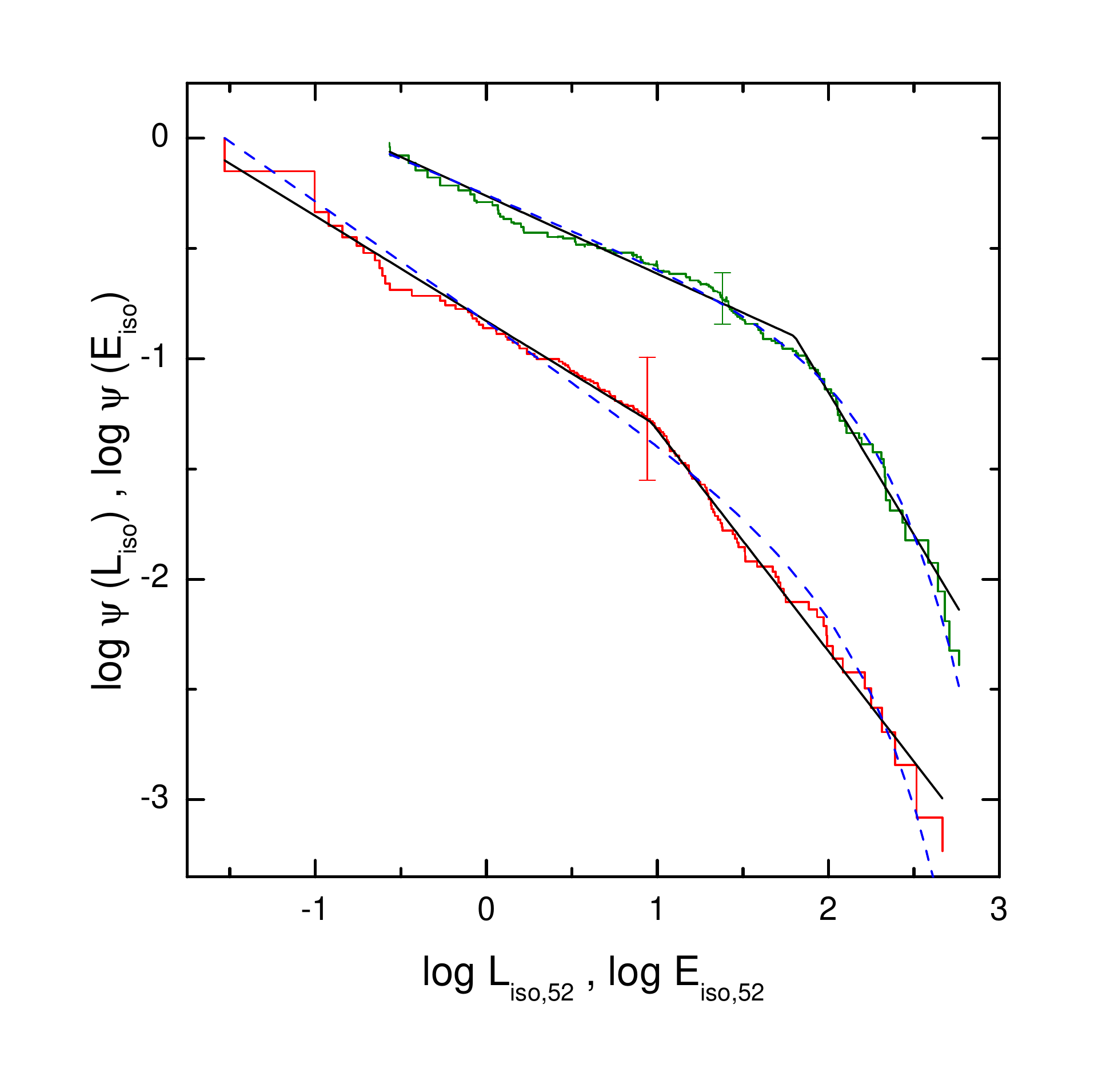}
\includegraphics[width=0.49\textwidth]{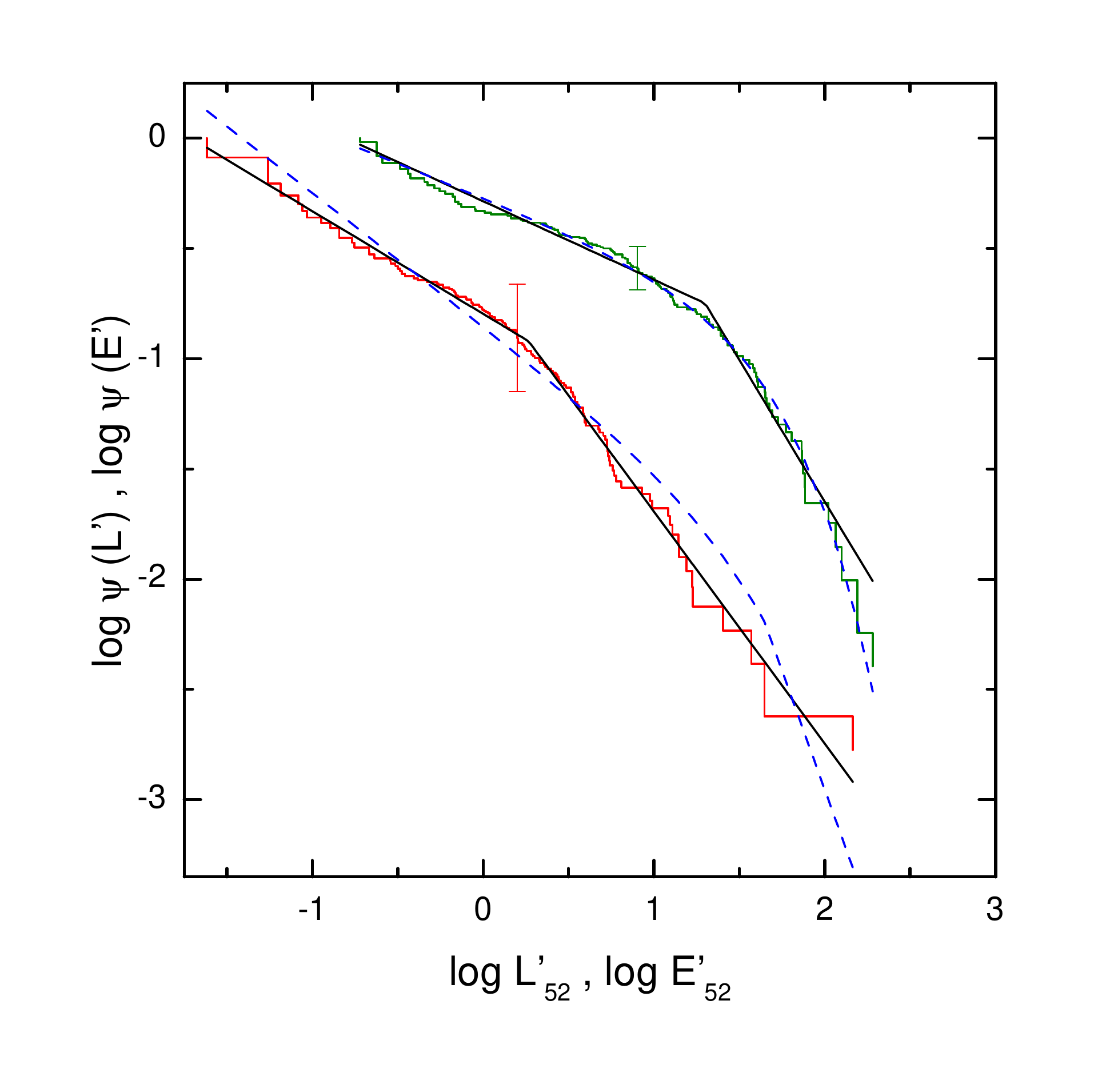}
\caption{Cumulative GRB isotropic-luminosity and isotropic-energy functions. Left panel: LF (red stepped graph) and EF (green stepped graph) estimated under the assumption of no evolution of $L_\textrm{iso}$ and $E_\textrm{iso}$ with $z$; the solid and dashed lines show the best BPL and CPL fits, respectively.
Right panel: present-time LF and EF estimated accounting for the luminosity and energy evolutions: $L'=L_\textrm{iso}/(1+z)^{1.7}$, $E'=E_\textrm{iso}/(1+z)^{1.1}$.
The distributions are normalized to unity at the dimmest points and a typical error bar is shown for each distribution.
}
\label{Cumulative_distr_LE}
\end{figure}

The fits were performed in $\textrm{log}- \textrm{log}$ space using $\chi^2$ minimization, the results are given in Table~\ref{tabLFEF} and shown in Figure~\ref{Cumulative_distr_LE} (right panel).
The derived BPL slopes of LF and EF are close to each other, both for the dim and bright segments, thus the shape of EF is similar to that of LF;
also, these indices are roughly consistent with the LF and EF slopes obtained in \citet{Yonetoku2004} and \citet{Wu2012}.
The small reduced $\chi^2$ obtained for both models do not allow us to reject any of them;
however, when compared to BPL, the CPL fit to $\psi(L')$ results in a considerably worse quality ($\chi^2_\textrm{CPL}-\chi^2_\textrm{BPL}>16$);
with the PL slope $\alpha \sim -0.6$ and the cutoff luminosity $L'_\textrm{cut} \sim 2.3 \times 10^{54}$~erg~s$^{-1}$.
Conversely, the cutoff PL fits $\psi(E')$ better when compared to BPL ($\chi^2_\textrm{CPL}-\chi^2_\textrm{BPL} \sim -5.5$);
with $\alpha \sim -0.35$ and the cutoff energy $E'_\textrm{cut} \sim 2.3 \times 10^{54}$~erg.
The existence of a sharp cutoff of the isotropic energy distribution distribution of KW and \emph{Fermi}/GBM GRBs around $\sim 1-–3 \times 10^{54}$~erg was suggested recently by \cite{Atteia2017}.

The derived $\psi(L^\prime)$ and $\psi(E^\prime)$ correspond to the present-time GRB luminosity and energy distributions (at $z=0$) and hence the local LF and EF in the comoving frame are roughly estimated as $\psi(L^\prime) (1+z)^{\delta_\textrm{L}}$ and $\psi(E^\prime) (1+z)^{\delta_\textrm{E}}$, correspondingly.
Taking into account that the $z$--$L_\textrm{iso}$ and $z$--$E_\textrm{iso}$ evolutions are established at $<2\sigma$,
the LF and EF calculated without accounting for the evolution, $\psi(L_\textrm{iso})$ and $\psi(E_\textrm{iso})$, may be of interest.
We estimated these functions by setting  $\delta_L$ and $\delta_E$ to zero, and found them very similar in shape to the present-time LF and EF (Figure~\ref{Cumulative_distr_LE}).
The results of the BPL and CPL fits to $\psi(L_\textrm{iso})$ and $\psi(E_\textrm{iso})$ are given in the last four lines of Table~\ref{tabLFEF}.

Finally, using the EP method, we estimated the cumulative GRB number distribution $\psi (z)$ and the derived GRBFR per unit time per unit comoving volume $\rho(z)$ (see Appendix for the details).
In Figure~\ref{GRBFR} we compare the star formation rate (SFR) data from the literature (\citealt{Hopkins2004}, \citealt{Bouwens2011}, \citealt{Hanish2006}, \citealt{Thompson2006}, and \citealt{Li2008})
with GRBFRs derived from different $z$--$L$ and $z$--$E$ distributions.
The GRBFR estimated from the evolution-corrected $z$--$L'$ distribution exceeds the SFR at $z<1$ and nearly traces the SFR at higher redshifts; 
the same behavior is noted for the GRBFRs estimated using both the evolution-corrected  $z$--$E'$ and the non-corrected $z$--$E_\textrm{iso}$ distributions.
The low-$z$ GRBFR excess over SFR is in agreement with the results reported in \citet{Yu2015} and \citet{Petrosian2015}.
Meanwhile, the only GRBFR that traces the SFR in the whole KW GRB redshift range is the $\rho(z)$ derived from the $z$--$L_\textrm{iso}$ distribution (i.e. not accounting for the luminosity evolution). 
Such behavior is known e.g. from \cite{Wu2012}, albeit for the GRBFR estimated from the $z$--$E_\textrm{iso}$ distribution.

\subsection{Hardness-duration distribution in the observer and rest frames}
\label{Discussion_hardness}
\begin{figure}
\center
\includegraphics[width=0.8\textwidth]{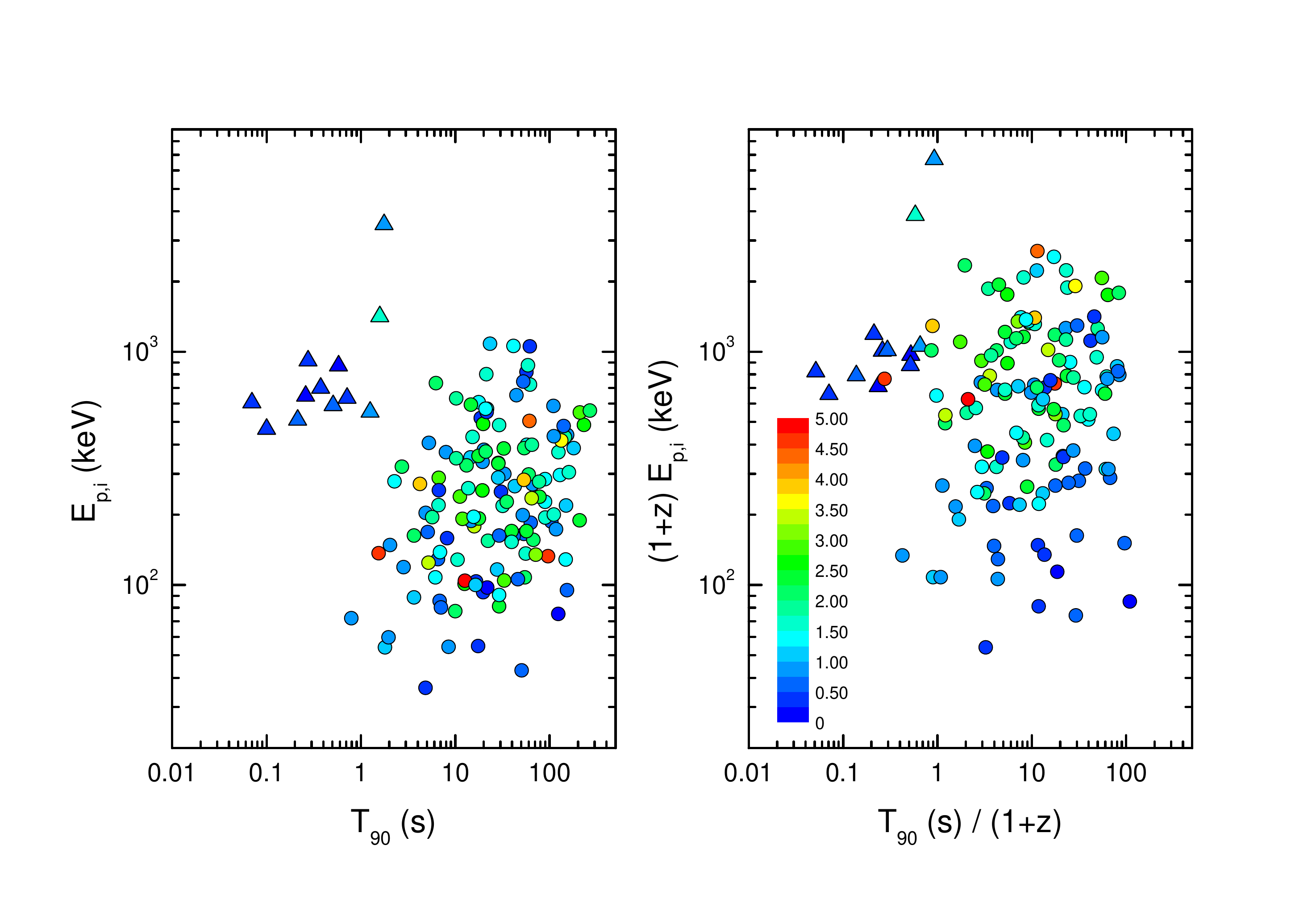}
\caption{$E_\textrm{p,i}$--$T_{90}$ diagram in the observer (left panel) and rest (right panel) frames.
The color of each data point (Type~I: triangles, Type~II: circles) represents the burst redshift.}
\label{GraphEpiT90}
\end{figure}

Figure~\ref{GraphEpiT90} shows $E_\textrm{p,i}$ as a function of the burst durations $T_{90}$ in the observer and rest frames.
In the observer frame the KW Type~I GRBs are typically harder and shorter than Type~II bursts, which is consistent with the classification
obtained from the hardness-duration distribution (Figure~\ref{GrHRT50}), and this tendency shows no dependence on the burst redshift.


In the cosmological rest frame this pattern remains practically unchanged for GRBs at $z \lesssim 1.7$
but it appears to be less distinct when the whole sample is considered.
Although in the rest frame Type~I GRBs are still shorter than Type~II GRBs, their rest-frame $E_\textrm{p}$, clustered around 1~MeV,
are superseded by those of a significant fraction of the Type~II population. The notable exceptions here are GRB~090510 and GRB~160410A,
whose rest-frame peak energies exceed those of even the highest-z Type~II GRBs.
We note, however, that the derived rest-frame durations are affected by a variable energy-dependant factor (Section~\ref{Temporal})
and the KW rest-frame $E_\textrm{p}$ are subject to the observational bias (Section~\ref{Selection}) thus an interpretation of the rest-frame
hardness-duration distribution should be done with care.

\subsection{Hardness-intensity correlations}
\label{Correlations}
Using the data described in the previous sections, we tested KW GRB characteristics against $E_\textrm{p,i}$--$S$ and $E_\textrm{p,p}$--$F_\textrm{peak}$ correlations in the observer frame,
and $E_\textrm{p,i,z}$--$E_\textrm{iso}$ (``Amati'') and $E_\textrm{p,p,z}$--$L_\textrm{iso}$ (``Yonetoku'') correlations in the rest frame,
along with their collimated versions $E_\textrm{p,i,z}$--$E_\gamma$ and $E_\textrm{p,p,z}$--$L_\gamma$.

To probe the existence of correlations, we calculated the Spearman rank-order correlation coefficients ($\rho_S$) and the associated null-hypothesis (chance) probabilities or p-values ($P_{\rho_S}$; \citealt{Press1992}). The null hypothesis is that no correlation exists; therefore, a small p-value indicates a significant correlation.
It was shown that the Nukers' estimate is an unbiased slope estimator for the linear regression \citep{Tremaine2002}. The Nukers' estimate is based on minimizing:
$$\chi^2 = \sum_{i=1}^{N} \frac{(y_i-a x_i-b)^2}{a^2 \sigma^2_{x i}+\sigma^2_{y i}+\sigma^2_\textrm{int}},$$
where $\sigma^2_{x i}$ and $\sigma^2_{y i}$ are the measurement errors; thus both variables are treated symmetrically in terms of their errors
and there is no need to choose dependent and independent variables.
Although a correlation may be highly significant, the reduced statistic, $\chi^2_r$, may be $\gg1$ indicating that either the uncertainties are underestimated or there is an intrinsic dispersion in the correlation. To account for the intrinsic dispersion, an additional term, $\sigma^2_\textrm{int}$, is added to the denominator and, in this case, $\chi^2_r$ is adjusted to ensure $\chi^2_r=1$.
Therefore, we approximated a linear regression between $\textrm{log}$-energy and $\textrm{log}$~$E_\textrm{p}$ using two methods, without $\sigma_\textrm{int}$ and with the intrinsic scatter.


Table~\ref{correlationtab} summarizes the correlation parameters we obtained for subsamples of Type~I GRBs, Type~II GRBs, and Type~II GRBs with $t_\textrm{jet}$ estimates.
The first column presents the correlation. The next three columns provide the number of bursts in the fit sample, $\rho_S$, and $P_{\rho_S}$.
The next columns specify the slopes ($a$), the intercepts ($b$), and $\sigma_\textrm{int}$.
Since zeroing the intrinsic scatter yields $\chi^2_r \gg 1$ for all the subsamples (and that confirms the relevance of accounting for $\sigma_\textrm{int}$),
their values are of little interest and we do not present the fit statistics in the Table.

For the subsamples of Type~I and Type~II KW GRBs both the Amati and Yonetoku correlations improve considerably when moving from the observer frame to the GRB rest frame ($\Delta \rho_S \geq$0.1),
with only marginal changes in the slopes.
We found the rest-frame correlations for Type~II bursts to be the most significant, with $P_{\rho_S}<2 \times 10^{-21}$.
The derived slopes of the Amati and Yonetoku relations for those GRBs are very close to each other, 0.469 ($\rho_S$=0.70, 138 GRBs) and 0.494 ($\rho_S$=0.73, 137 GRBs), respectively.
These values are in agreement with the original results of \cite{Amati2002} and \cite{Yonetoku2004} and their further improvements (e.g. \citealt{Nava2012}).
When accounting for the intrinsic scatter, these slopes change to a more gentle $\sim$0.35 (with $\sigma_\textrm{int}\sim$0.24).

\begin{figure}
\center
\includegraphics[width=0.4\textwidth]{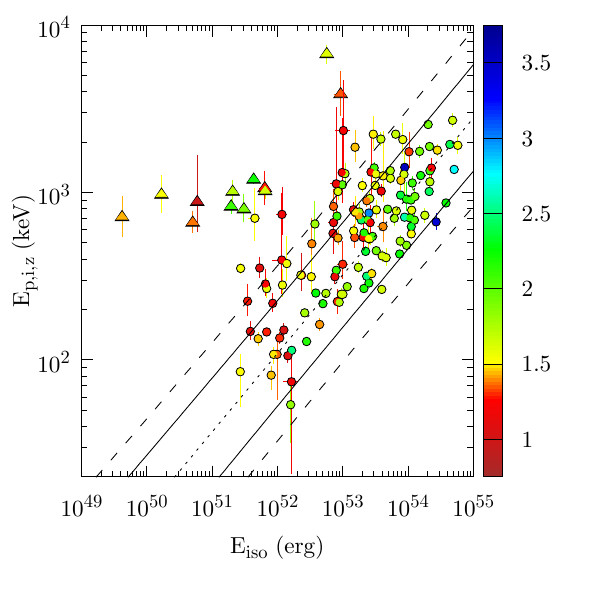}
\includegraphics[width=0.4\textwidth]{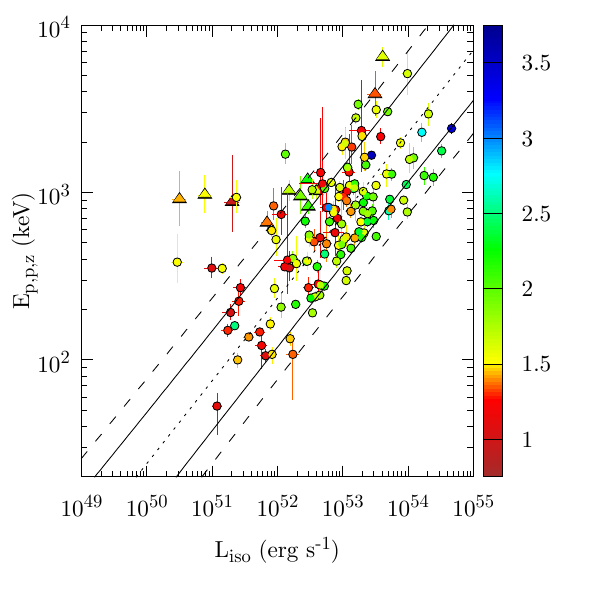}
\includegraphics[width=0.4\textwidth]{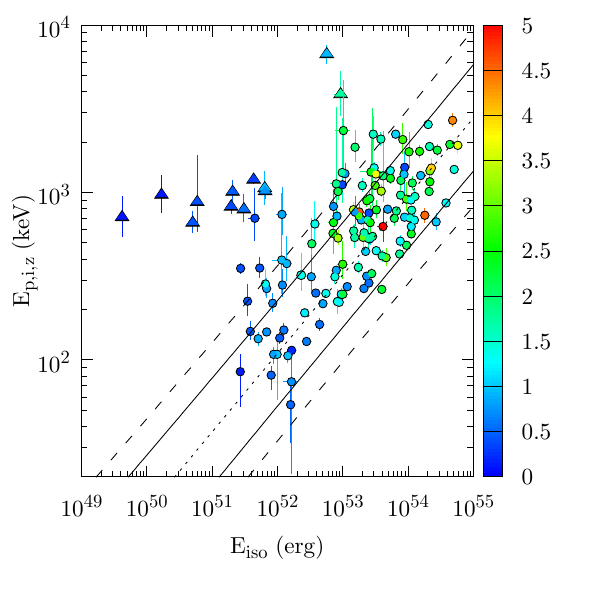}
\includegraphics[width=0.4\textwidth]{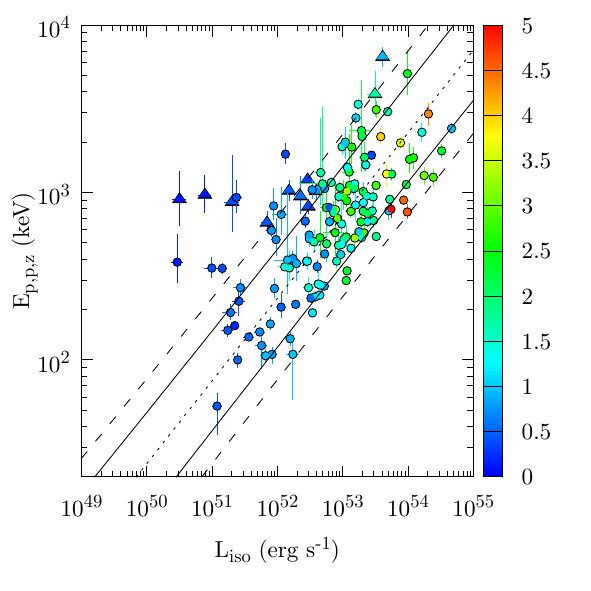}
\includegraphics[width=0.4\textwidth]{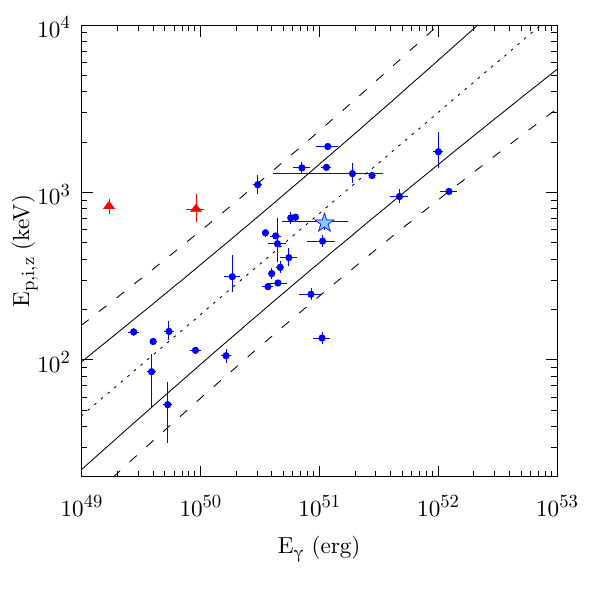}
\includegraphics[width=0.4\textwidth]{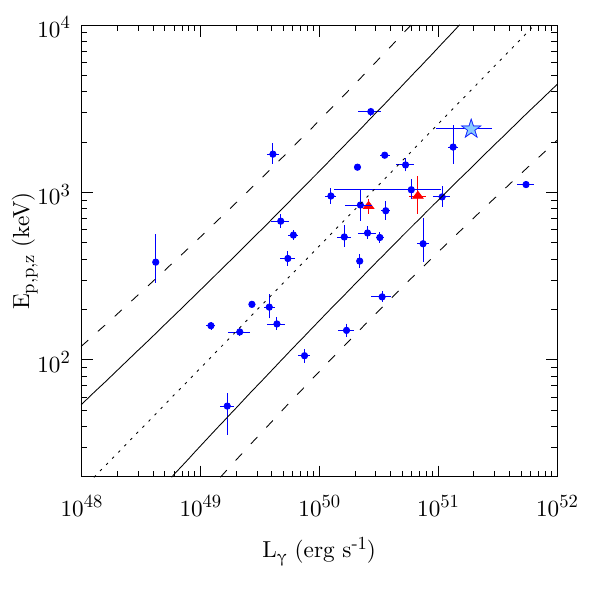}
\caption{Rest-frame energetics in the $E_\textrm{iso}–-E_\textrm{p,i,z}$ (left) and $L_\textrm{iso}–-E_\textrm{p,p,z}$ (right) planes.
The color of each data point (Type I: triangles, Type II: circles) represents the log of the burst's trigger significance (upper panels)
and the GRB redshift (middle panels).
The ``Amati'' and ``Yonetoku'' relations calculated without internal scattering for Type II GRBs are plotted with dotted lines;
the solid and dashed lines show their 68\% and 90\% PI's, respectively.
The lower panels present these relations for collimation-corrected energetics where the ultraluminous GRB~110918A is shown with the star.
}
\label{AmatiYonetoku}
\end{figure}

As one can see in Figure~\ref{AmatiYonetoku}, the lower boundaries of both the Amati and Yonetoku relations are defined by GRBs with moderate-to-high detection significance, so the instrumental biases do not affect the correlations from this edge of the distributions.
Meanwhile, all outliers in the relations lie \emph{above} the upper boundaries of the 90\% prediction intervals (PIs) of the relations.
Since these bursts were detected at lower significance, with the increased number of GRB redshift observations, one could expect a ``smear'' of the hardness-intensity correlations due to more hard-spectrum/less-energetic GRB detections.
Thus, using the KW sample, we confirm a finding of \cite{Heussaff2013} that the lower right boundary of the Amati correlation (the lack of luminous soft GRBs) is an intrinsic GRB property, while the top left boundary may be due to selection effects. This conclusion may also be extended to the Yonetoku correlation.

The collimated versions of these relations were tested on the subsample of 30 Type~II GRBs with reliable $t_\textrm{jet}$ (last four lines of Table~\ref{correlationtab}).
We found that accounting for the jet collimation for the KW sample neither improves the significance of the correlations nor reduces the dispersion of the points around the best-fit relations.
The slopes we obtained for the collimated Amati and Yonetoku relations are steeper compared to those of the non-collimated versions.

The $E_\textrm{p,i,z}$--$E_\textrm{iso}$ and $E_\textrm{p,p,z}$--$L_\textrm{iso}$ correlations for 12 Type~I bursts are less significant when compared to those for Type~II GRBs,
and they are characterized by less steep slopes (0.364 and 0.396 for $E_\textrm{p,i,z}$--$E_\textrm{iso}$ and $E_\textrm{p,p,z}$--$L_\textrm{iso}$, respectively).
It should be noted, however, that the rest-frame $E_\textrm{p,i}$ of Type~I GRBs shows only a weak (if any) dependence on the burst energy below $E_\textrm{iso}\sim 10^{52}$~erg
(Figure~\ref{AmatiYonetoku}), and the same is true for the  $E_\textrm{p,p,z}$--$L_\textrm{iso}$ relation at $L_\textrm{iso}\lesssim 5 \times 10^{52}$~erg/s.
Above these limits the slopes of both relations for Type~I GRBs are similar to those for Type~II GRBs.
As one can see from the Figure, all KW Type~I bursts are hard-spectrum/low-isotropic-energy outliers in the Amati relation for Type~II GRBs.
In the $E_\textrm{p,p,z}$--$L_\textrm{iso}$ plane this pattern is less distinct; at luminosities above $L_\textrm{iso}\sim 10^{52}$~erg/s the Type~I bursts nearly follow the upper boundary of the Type~II GRB Yonetoku relation. Finally, the two KW Type~I GRBs with available collimation data lie above 90\% PI for the Type~II GRB $E_\textrm{p,i,z} - E_\gamma$ relation and, simultaneously, within the 68\% PI for the $E_\textrm{p,p,z}$--$L_\gamma$ relation (Figure~\ref{AmatiYonetoku}, lower panels).

We also calculated the collimation-corrected energetics for the ultraluminous KW GRB~110918A using $t_\textrm{jet}=0.2\pm0.13$~days estimated by \cite{Frederiks2013} from an extrapolation of early $\gamma$-ray/late X-ray afterglow data. As can be seen in Figure~\ref{AmatiYonetoku}, the implied $E_\gamma \approx 1.1 \times 10^{51}$~erg and $L_\gamma \approx 1.9 \times 10^{51}$~erg~s$^{-1}$ nicely agree with both hardness-intensity relations for our ``reliable $t_\textrm{jet}$'' GRB sample. This supports the correctness of the $t_\textrm{jet}$ estimate and favors
the conclusion of \cite{Frederiks2013} that a tight collimation of the jet ($\theta_\textrm{jet} \sim 1.6 \arcdeg$) must have been a key ingredient to produce this unusually bright burst.

\section{SUMMARY AND CONCLUSIONS}
\label{Conclusions}
We have presented the results of a systematic study of 150 GRBs with reliable redshift estimates detected in the triggered mode of the Konus-Wind experiment.
The sample covers the period from 1997 February to 2016 June and represents the largest set of cosmological GRBs to date over a broad energy band.
Among these GRBs, twelve bursts (or 8\%) belong to the Type I (merger origin, short/hard) GRB population and the others are Type II (collapsar origin, long/soft) bursts.

From the temporal and spectral analyses of the sample, we provide the burst durations $T_{100}$, $T_{90}$, and $T_{50}$, the spectral lags, and spectral fits with CPL and Band model functions.
From the BEST spectral models we calculated the 10~keV--10~MeV energy fluences and the peak energy fluxes on three time scales, including the GRB rest-frame 64~ms scale.
Based on the GRB redshifts, which span the range $0.1 \leq z \leq 5$, we estimated the rest-frame, isotropic-equivalent energies ($E_\textrm{iso}$) and peak luminosities ($L_\textrm{iso}$).
For 32 GRBs with reasonably constrained jet breaks we provide the collimation-corrected values of the energetics.

We analyzed the influence of instrumental selection effects on the GRB parameter distributions and found that the regions above the limits,
corresponding to the bolometric fluence $S_\textrm{lim} \sim 3-4 \times 10^{-6}$~erg cm$^{-2}$ (in the $E_\textrm{iso}$--$z$ plane)
and bolometric peak energy flux $F_\textrm{lim} \sim 1-2 \times 10^{-6}$~erg~cm$^{-2}$~s$^{-1}$ (in the $L_\textrm{iso}$--$z$ plane) may be considered free from selection biases.
For the bursts in our sample we calculated the KW GRB detection horizon, $z_\textrm{max}$, which extends to $z \sim 16.6$, stressing the importance of GRBs as probes of the early Universe.
Among the KW short/hard GRBs the highest $z_\textrm{max}$ is $\approx 5.3$.

Accounting for the instrumental biases and using the non-parametric methods of \citet{Lynden-Bell1971} and EP, we estimated the GRB luminosity evolution,
luminosity and isotropic-energy functions, and the evolution of the GRB formation rate. The derived luminosity evolution and isotropic energy evolution indices $\delta_L\sim1.7$ and $\delta_E\sim1.1$ are more shallow than those reported in previous studies, albeit within errors. 
The shape of the derived LF is best described by the broken PL function with low- and high-luminosity slopes $\sim-0.5$ and $\sim-1$, respectively. 
The EF is better described by the exponentially-cutoff PL with the PL index $\sim-0.3$ and a cutoff isotropic energy of $\sim (2-4) \times 10^{54}$~erg. 
The derived GRBFR features an excess over the SFR at $z<1$ and nearly traces the SFR at higher redshifts.

We considered the behavior of the rest-frame GRB parameters in the hardness-duration and hardness-intensity planes, and confirmed the ``Amati'' and ``Yonetoku'' relations for Type II GRBs. We found that the correction for the jet collimation does not improve these correlations for the KW sample.
Using the KW sample, we confirm a finding of \cite{Heussaff2013} that the lower right boundary of the Amati correlation (the lack of luminous soft GRBs) is an intrinsic GRB property, while the top left boundary may be due to selection effects. This conclusion may also be extended to the Yonetoku correlation.

Plots of the the GRB light curves and spectral fits can be found at the Ioffe Web site\footnote{http://www.ioffe.ru/LEA/zGRBs/triggered/}.
We hope this catalog will encourage further investigations of GRB physical properties and will contribute to other related studies.

The authors are grateful to the anonymous referee for careful reading and constructive comments which improved the manuscript.
We thank Maria Giovanna Dainotti for a stimulating discussion and Vah\'{e} Petrosian for helpful comments.
This work was supported by RSF (grant 17-12-01378).
We acknowledge the use of the public data from the \textit{Swift} data archive\footnote{http://swift.gsfc.nasa.gov} and the use of the data from the Gamma-Ray Burst Online Index (``GRBOX'')\footnote{http://www.astro.caltech.edu/grbox/grbox.php}.

\textit{Facility:} \facility{\textit{Wind}(Konus)}

\clearpage
\begin{appendix}
\section{APPENDIX: Non-parametric statistical techniques for a truncated data sample}
\label{Appendix}
Here, we describe the details of the the non-parametric statistical techniques used to obtain the unbiased parameter distributions for a sample subject to selection effects in the $z$--$L_\textrm{iso}$ plane implying that the same methodology can be applied to the $z$--$E_\textrm{iso}$ plane.

\begin{figure}
\center
\includegraphics[width=\textwidth]{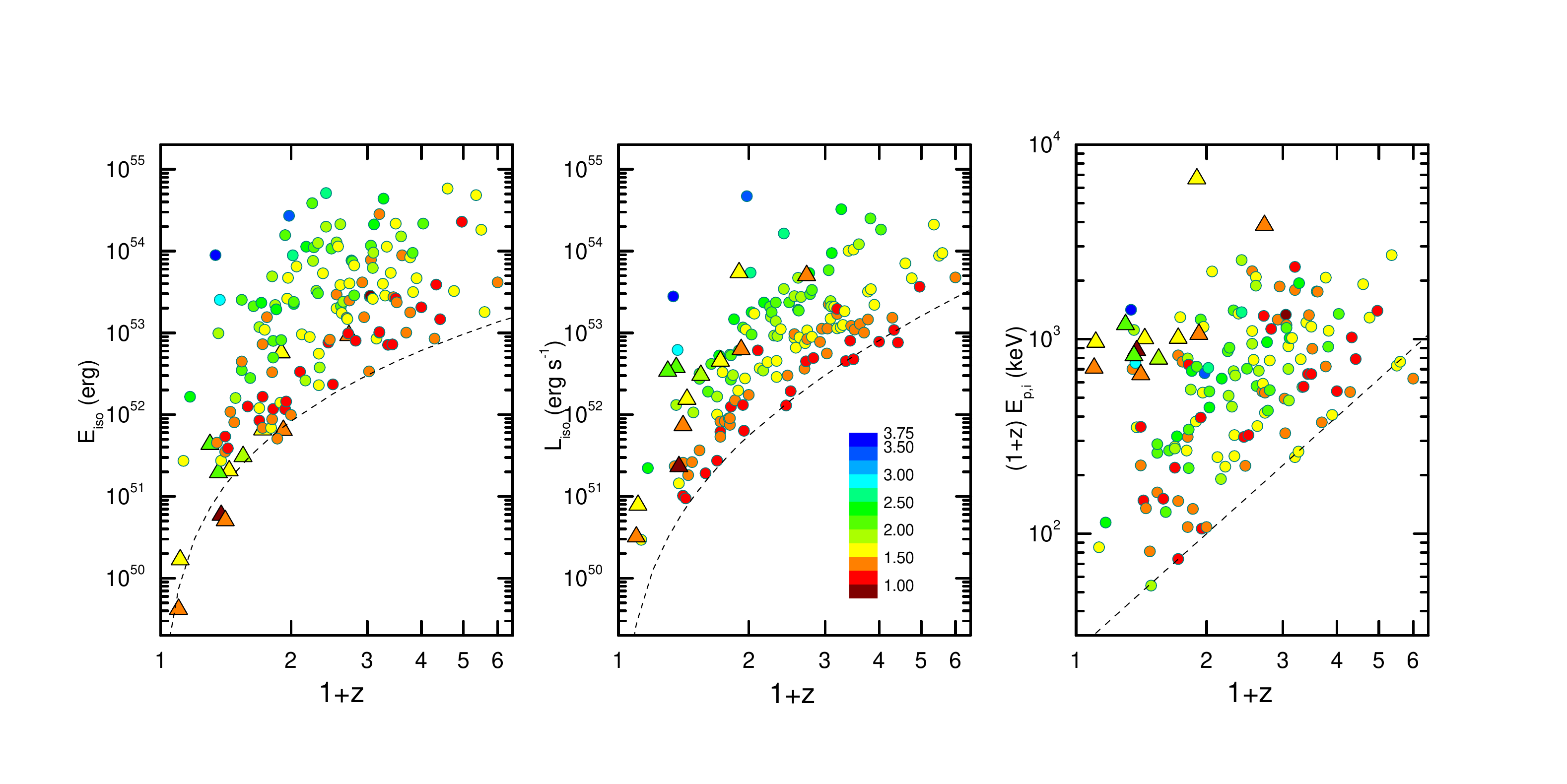}
\caption{KW GRB $E_\textrm{iso}$, $L_\textrm{iso}$, and $E_\textrm{p,i,z}$ vs. redshift.
The color of each data point (Type~I: triangles, Type~II: circles) represents the log of the burst's trigger significance ($\sigma$).
The observer-frame limits (Section~\ref{Selection}) are shown with dashed lines.}
\label{GraphSensitivity2}
\end{figure}

The $z$--$L_\textrm{iso}$ sample suffers from selection effect due to the detection limit of the instrument (see Section~\ref{Selection} for details), which results in the data truncation seen in Figure~\ref{GraphSensitivity2}. Although it is a common practice to estimate the trigger sensitivity as a ``characteristic'' energy flux that could trigger a detector,
the trigger threshold flux can actually depend on some parameters, e.g., the burst spectral shape, the background count rate,
the incident angle, and the calibration; the $k$-corrected flux also depends on the redshift.
Therefore, while deriving LF and GRBFR from the KW data we used the individual $k$-corrected trigger threshold fluxes $F_\textrm{lim}$
(see Section~\ref{Selection}) as a proxy for the instrumental selection effect.
The results obtained using a ``monolithic'' truncation curve, however, are very similar to those obtained with the first method.

The parent distributions can be obtained from the biased $z$--$L_\textrm{iso}$ sample using the non-parametric Lynden-Bell $C^-$ techniques (\citealt{Lynden-Bell1971}) further advanced by \citet{Efron1992}.
Moreover, as shown by \citet{Petrosian1992}, all nonparametric methods for determining the underlying distributions reduce to the \citet{Lynden-Bell1971} method in case of a one-sided truncation.
Initially developed for a truncated QSO sample, this procedure was first applied to the truncated GRB data by \citet{Lloyd-Ronning2002}.

Since the Lynden-Bell $C^-$ approach is applicable only if the luminosity and redshift distributions are independent, the dependence of $L$ on $z$ should be tested and rejected (if present).
For this purpose one can use the methodology developed by \citet{Efron1992}.
The EP method uses a modified version of the Kendall rank correlation coefficient (the Kendall $\tau$ statistic) to test the independence of variables in truncated data.
Instead of calculating the ranks of each data points among all observed objects, which is normally done for untruncated data,
the rank of each data point is determined among its ``associated set'' which include all objects that could have been observed given the observational limits.

\begin{figure}
\center
\includegraphics[width=0.8\textwidth]{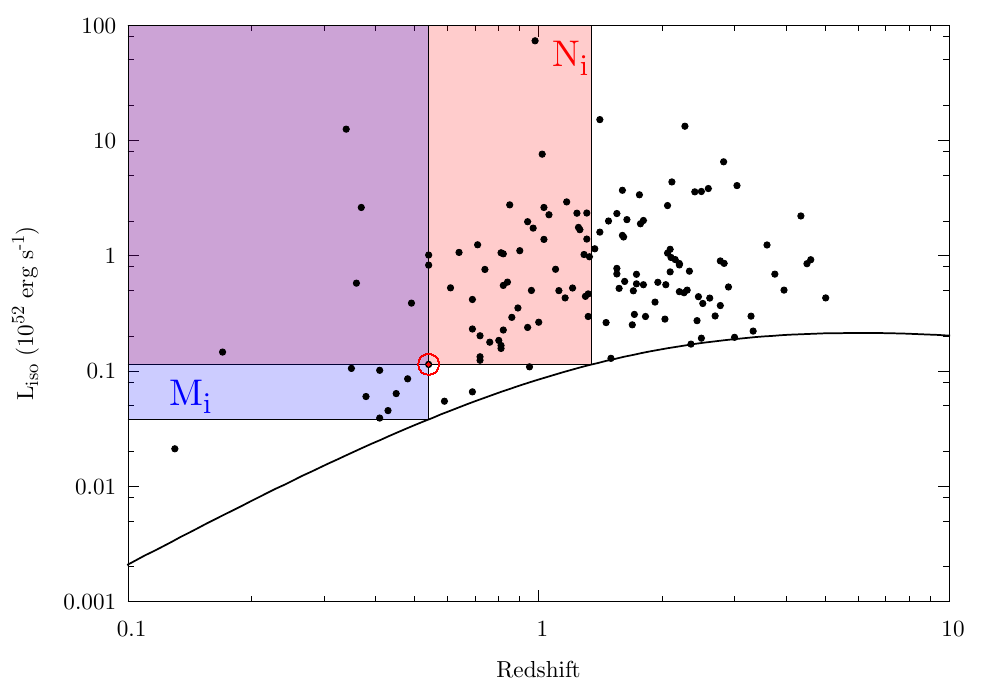}
\caption{Example of the associated set for the non-evolving luminosity sample. The line represents the truncation limit corrected for the $L_\textrm{iso}$ evolution. The``$N_i$''  and ``$M_i$'' denote the LF and GRBFR associated sets for the $i$th burst, correspondingly. See text for details.}
\label{Associated_set}
\end{figure}

Consider a set of observables $L_i$ and $z_i$, where $i$ is the burst index.
For each burst from the sample we construct an associated set of
$$J_i = \{j|L_j>L_i, L_i>L_\textrm{lim,j}\},$$
where $L_i$ is the $i$th GRB luminosity,
and $L_\textrm{lim,j}$ is the minimum observable luminosity at $z_j$.
Another commonly used definition of the associated set is $$J_i = \{j|L_j>L_i, z_j<z_\textrm{lim,i}\},$$
where $z_\textrm{lim,i}$ is the maximum redshift at which a GRB with luminosity $L_i$ can be observed,
and produces the same subsample of bursts as the foregoing definition if the truncation effect is a monotonic function.
An example of the associated set for the $i$th burst is shown in Figure~\ref{Associated_set}.

Let $N_i$ be the number of bursts in the $i$th associated set (that is the same as $C^-$ in \citealt{Lynden-Bell1971}) and
$R_i$ the number of events that have redshift $z_j$ less than $z_i$ (that is an analog of the $i$th burst rank in the associated set):
$$N_i = \textrm{Number}\{J_i\},$$
$$R_i = \textrm{Number}\{j \in J_i:z_j<z_i\}.$$
Then the degree of correlation between $L$ and $z$ can be estimated via the test statistic $\tau$ parametrized as
$$\tau = \frac{\sum_i(R_i - E_i)}{\sqrt{\sum_i V_i}},$$
where $E_i = (N_i+1)/2$ is the expected mean, and $V_i = (N_i^2-1)/12$ is the variance of the uniform distribution.
In the non-truncated case this $\tau$ statistic is equivalent to the Kendall's nonparametric correlation coefficient.
If $z_i$ and $L_i$ are independent of each other, then $R_i$ is uniformly distributed between 1 and $N_i$, therefore the samples $R_i\le E_i$ and $R_i \ge  E_i$ should be nearly equal, and the $\tau$ statistic will be close to 0.
Since the $\tau$ statistic is normalized by the square root of variance, the correlation coefficient between $z$ and $L$ is measured in units of the standard deviation.

\begin{figure}
\center
\includegraphics[width=0.50\textwidth]{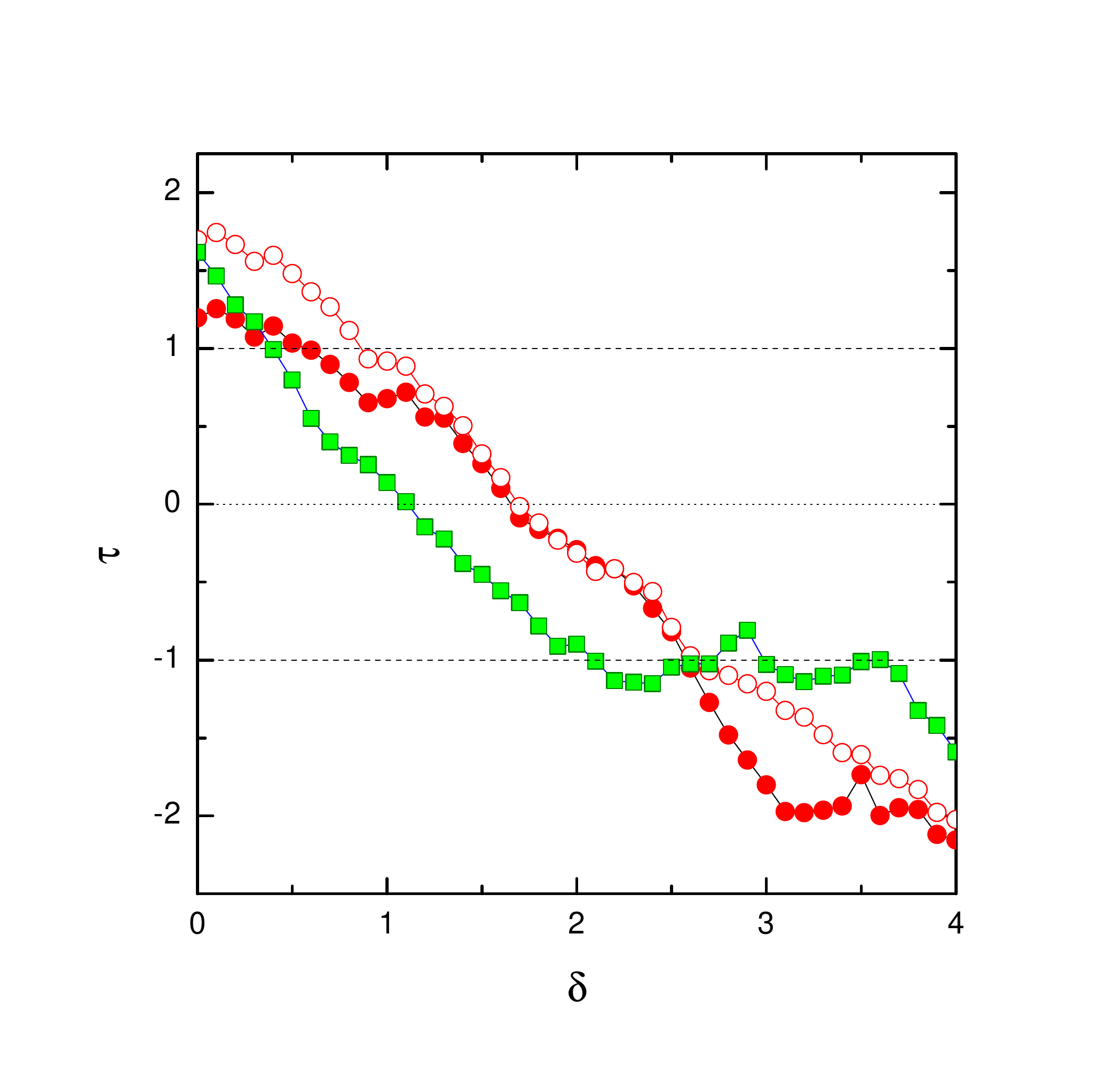}
\includegraphics[width=0.47\textwidth]{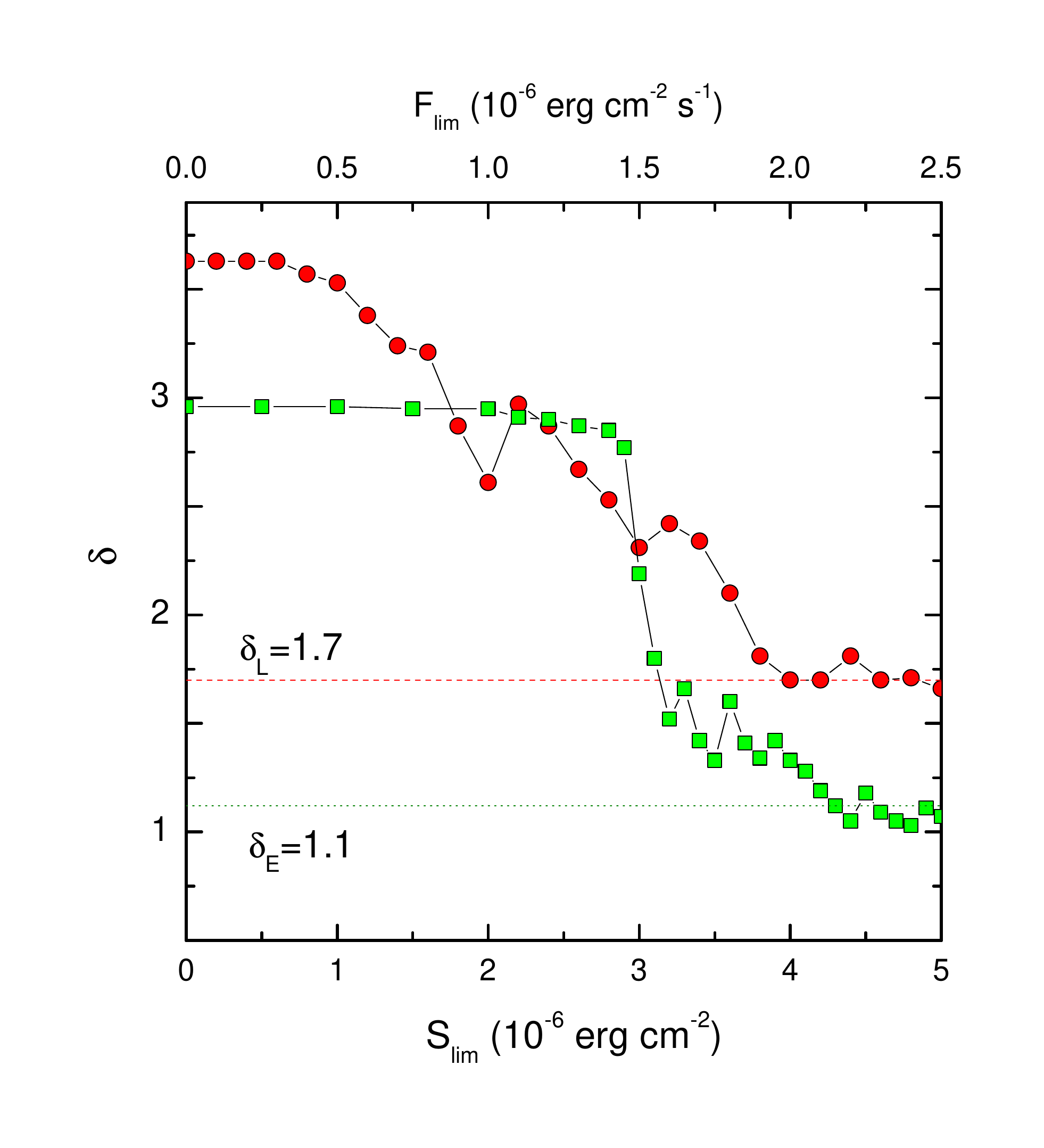}
\caption{Left: modified Kendall statistic $\tau$ vs. luminosity and isotropic-energy evolution indices $\delta_L$ (per-burst truncation flux $F_\textrm{lim}$, red open circles; monolithic $F_\textrm{lim} = 2 \times 10^{-6}$~erg~cm$^{-2}$~s$^{-1}$, red filled circles) and $\delta_E$ (green squares, monolithic $S_\textrm{lim} = 4.3 \times 10^{-6}$~erg~cm$^{-2}$). The values of $\delta$ for which $\tau = 0$ and $\tau = \pm 1$ give the best value and one sigma range for independence.
Right: dependency of the \textbf{best values of} $\delta_L$ and $\delta_E$ (red circles and green squares, respectively) on the monolithic \textbf{truncation limits} $F_\textrm{lim}$ and $S_\textrm{lim}$.
The dashed and dotted lines denote the ``settled'' values of $\delta_L$ and $\delta_E$, which correspond
to $F_\textrm{lim} \gtrsim 2 \times 10^{-6}$~erg~cm$^{-2}$~s$^{-1}$ and $S_\textrm{lim} \gtrsim 4.3 \times 10^{-6}$~erg~cm$^{-2}$, respectively.
}
\label{GrTau}
\end{figure}

Next, the index of the luminosity evolution $\delta$ should be varied to adjust the test statistic to $\tau(\delta) = 0$ for the luminosity $L^\prime = L/(1+z)^{\delta}$
and thus removing the effect of luminosity evolution. The $1 \sigma$ confidence interval on $\delta$ is obtained when $\tau = \pm 1$ (Figure~\ref{GrTau}, left panel) and the luminosity evolution is rejected at the $\tau_0\equiv\tau(\delta=0)$ level. In case the ``monolithic'' truncation curve is used, the resulting evolution index $\delta$ is strongly dependent on the limiting flux (or fluence).
We investigated the dependency of the luminosity and energy evolution indices $\delta_L$ and $\delta_E$ on the corresponding truncation limits $F_\textrm{lim}$ and $S_\textrm{lim}$
for the KW sample (Figure~\ref{GrTau}, right panel) and determined the limits $F_\textrm{lim} \gtrsim 2 \times 10^{-6}$~erg~cm$^{-2}$~s$^{-1}$ and $S_\textrm{lim} \gtrsim 4.3 \times 10^{-6}$~erg~cm$^{-2}$
above which $\delta_L$ and $\delta_E$ do not vary much with the truncation limit change and fluctuate around the ``settled'' values $\delta_L\sim1.7$ and $\delta_E\sim1.1$.
Interestingly, a similar value of $\delta_L$ ($\sim$1.7) is obtained when the individual truncation limits are used for each burst.

Once obtained, the luminosity evolution index $\delta_L$, the observed luminosity $L_\textrm{iso}$ can be converted into the local (non-evolving) luminosity space $L^\prime = L_\textrm{iso}/(1+z)^{\delta_L}$. Then, following \citet{Lynden-Bell1971}, the local cumulative LF $\psi(L^\prime)$can be non-parametrically derived as a function of univariate $L^\prime$:
$$\ln \psi(L^\prime_i) = \sum_{j=2}^i \ln\left(1+\frac{1}{N^\prime_j}\right),$$
where $N^\prime_j$ is the number of points in the $i$th associated set for the local luminosities.

To estimate the cosmic GRBFR from the $z$--$L^\prime$ sample, we produce a cumulative number distribution $\psi(z)$.
First, we generate an associated set $$J^\prime_i = \{j|z_j<z_i, L_j>L_\textrm{lim,i}, L_\textrm{i}>L_\textrm{lim,j}\}$$
with $M_i$ points in each associated set (see Figure~\ref{Associated_set} for an example of an associated set obtained for a truncation curve).
The condition $ L_j>L_\textrm{lim,i}$ can be expressed as $z_\textrm{lim,j} > z_i$, but the $z_\textrm{lim}$ estimation is complicated in case of a non-analytic truncation boundary.
In the case where we used a set of threshold luminosities instead of a monotonic truncation curve, we applied an additional criterion of $L_i>L_\textrm{lim,j}$ to ensure that all the bursts of the associated set are not being subject to selection effect.
Then we calculate the cumulative function
$$\ln \psi(z_i) = \sum_{j=2}^i \ln\left(1+\frac{1}{M_j}\right).$$
Since the differential form of the GRBFR is more useful for comparison with the SFR,
we convert $\psi(z_i)$ into a differential form:
$$\rho(z) = \frac{d \psi}{d z}(1+z)\left(\frac{dV(z)}{dz}\right)^{-1}$$
where the additional factor $(1+z)$ comes from the cosmological time dilation, required when measuring a rate, and $dV(z)/dz$ is the differential comoving volume:
$$\frac{dV(z)}{dz} = \frac{4 \pi D_\textrm{H} D_\textrm{M}^2}{E(z)},$$
where $D_\textrm{M}$ is the transverse comoving distance, $D_\textrm{H} = c/H_0$ is the Hubble distance, and $E(z) = \sqrt{\Omega_\textrm{M}(1+z)^3+\Omega_\Lambda}$ is the normalized Hubble parameter.

\end{appendix}

\clearpage
\bibliography{GRBs_with_redshifts}{}

\clearpage

\clearpage

\clearpage

\clearpage

\clearpage

\clearpage

\clearpage

\clearpage

\clearpage

\clearpage

\begin{figure}
\center
\includegraphics[width=0.8\textwidth]{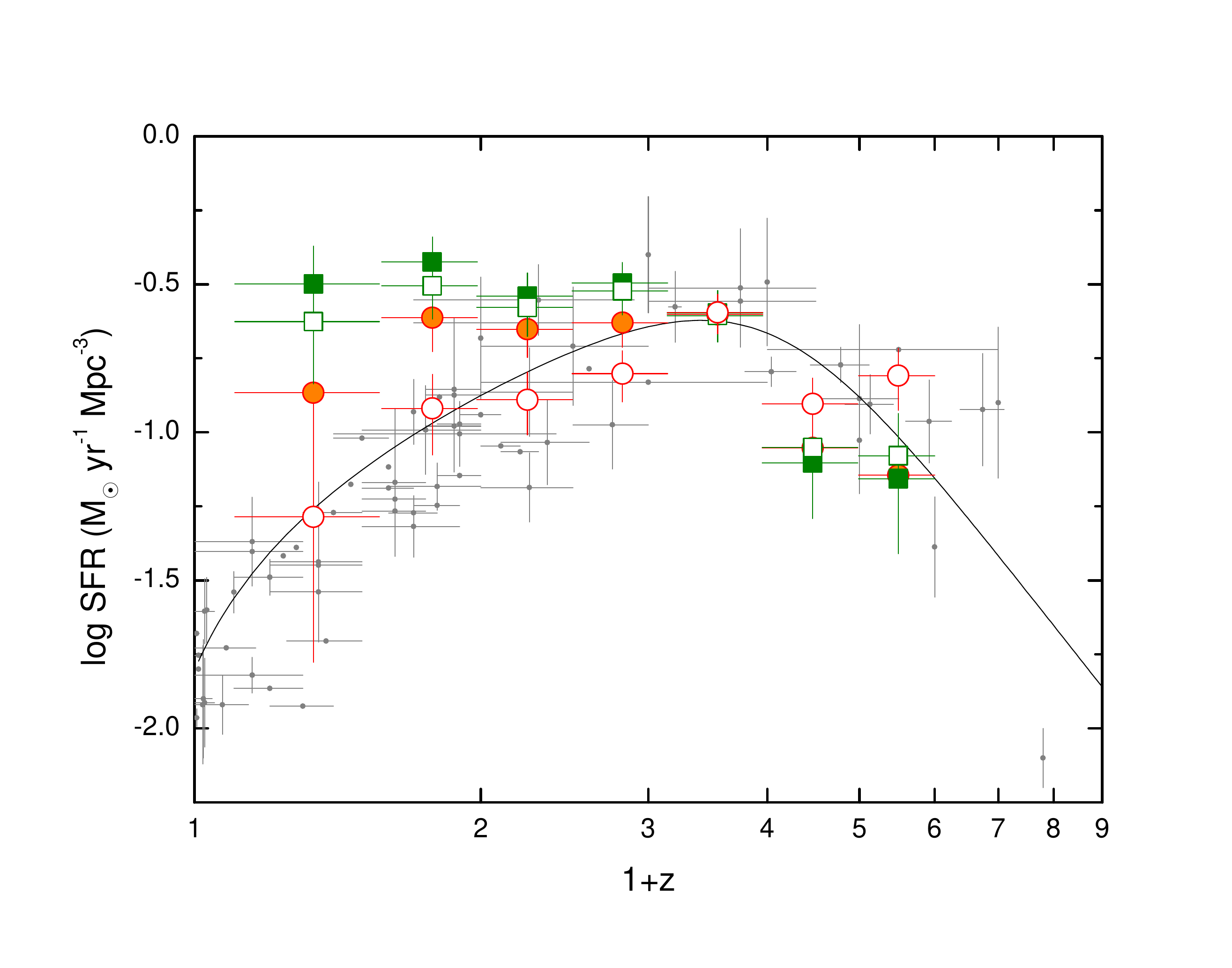}
\caption{Comparison of the derived GRBFR and the SFR data from the literature.
The GRBFR was calculated using four datasets: $z$--$L_\textrm{iso}$ (no luminosity evolution, red open circles), $z$--$L'$ ($\delta_L=1.7$, red filled circles), $z$--$E_\textrm{iso}$ (no energy evolution, green open squares), and $z$--$E'$ ($\delta_E=1.1$, green filled squares).
The gray points show the SFR data from \citet{Hopkins2004}, \citet{Bouwens2011}, \citet{Hanish2006}, and \citet{Thompson2006}.
The black solid line denotes the SFR approximation from \citet{Li2008}.
The GRBFR points have been shifted arbitrarily to match the SFR at $(1+z)\sim3.5$.}
\label{GRBFR}
\end{figure}

\clearpage

\clearpage

\clearpage
\begin{deluxetable}{lccccccccccc}
\rotate
\tabletypesize{\scriptsize}
\tablewidth{0pt}
\tablecaption{General information\label{generaltab}}
\tablehead{
\colhead{Burst}
& \colhead{Trigger}
& \colhead{Geocenter}
& \colhead{Type}
& \colhead{Local.}
& \colhead{Local.}
& \colhead{Other \tablenotemark{a}}
& \colhead{Det.}
& \colhead{Inc. angle}
& \colhead{$z$}
& \colhead{$z$\ type\tablenotemark{b}}
& \colhead{$z$}
\\
\colhead{name}
& \colhead{time}
& \colhead{time}
& \colhead{}
& \colhead{instr.}
& \colhead{ref}
& \colhead{obs.}
& \colhead{}
& \colhead{(deg)}
& \colhead{}
& \colhead{}
& \colhead{ref}
}
\startdata
GRB 970228 & 02:58:01.317 & 02:58:01.263 & II & \textit{BeppoSAX} & (\ref{Gen:BeppoSAXcat}) &  3 & S1 & 79.1 & $0.695$ & s & (\ref{Gen:Bloom2001}) \\
GRB 970828 & 17:44:42.357 & 17:44:42.157 & II & RXTEASM & (\ref{Gen:Remillard1997}) &  & S2 & 7.1 & $0.9578$\tablenotemark{\dag} & s & (\ref{Gen:Djorgovski2001}) \\
GRB 971214 & 23:20:52.214 & 23:20:51.043 & II & \textit{BeppoSAX} & (\ref{Gen:BeppoSAXcat}) & 1 3 & S2 & 33.9 & $3.418$ & s & (\ref{Gen:Kulkarni1998}) \\
GRB 990123 & 09:47:14.151 & 09:47:15.040 & II & \textit{BeppoSAX} & (\ref{Gen:BeppoSAXcat}) & 1 3 & S2 & 29.9 & $1.6004$ & s & (\ref{Gen:Kulkarni1999}) \\
GRB 990506 & 11:23:30.813 & 11:23:31.274 & II & \textit{BeppoSAX} & (\ref{Gen:BeppoSAXcat}) & 1 3 & S1 & 65.1 & $1.30658$\tablenotemark{\dag} & s & (\ref{Gen:Bloom2003a}) \\
GRB 990510 & 08:49:10.059 & 08:49:10.052 & II & \textit{BeppoSAX} & (\ref{Gen:BeppoSAXcat}) & 1 3 & S1 & 28.7 & $1.6187$ & s & (\ref{Gen:Vreeswijk2001}) \\
GRB 990705 & 16:01:26.864 & 16:01:26.352 & II & \textit{BeppoSAX} & (\ref{Gen:BeppoSAXcat}) &  3 & S1 & 7.1 & $0.8424$ & s & (\ref{Gen:LeFloch2002}) \\
GRB 990712 & 16:43:06.123 & 16:43:03.763 & II & \textit{BeppoSAX} & (\ref{Gen:BeppoSAXcat}) &  3 & S1 & 33.2 & $0.4331$ & s & (\ref{Gen:Vreeswijk2001}) \\
\enddata
\end{deluxetable}

\clearpage
\begin{deluxetable}{lrrr@{$\pm$}lr@{$\pm$}lr@{$\pm$}lr@{$\pm$}lr@{$\pm$}lr@{$\pm$}lr@{$\pm$}lr@{$\pm$}l}
\tabletypesize{\tiny}
\tablewidth{0pt}
\tablecaption{Durations and spectral lags\label{temporaltab}}
\tablehead{
\colhead{Burst}
& \colhead{$t_{0}$}
& \colhead{$T_{100}$}
& \multicolumn{2}{c}{$t_{5}$}
& \multicolumn{2}{c}{$T_{90}$}
& \multicolumn{2}{c}{$t_{25}$}
& \multicolumn{2}{c}{$T_{50}$}
& \multicolumn{2}{c}{$\tau_\textrm{lagG2G1}$}
& \multicolumn{2}{c}{$\tau_\textrm{lagG3G1}$}
& \multicolumn{2}{c}{$\tau_\textrm{lagG3G2}$}
\\
\colhead{Name}
& \colhead{(s)}
& \colhead{(s)}
& \multicolumn{2}{c}{(s)}
& \multicolumn{2}{c}{(s)}
& \multicolumn{2}{c}{(s)}
& \multicolumn{2}{c}{(s)}
& \multicolumn{2}{c}{(s)}
& \multicolumn{2}{c}{(s)}
& \multicolumn{2}{c}{(s)}
}
\startdata
GRB 970228 & $-0.456$ & $56.584$ & $-0.002$ & $0.024$ & $53.442$ & $2.891$ & $0.592$ & $0.048$ & $39.280$ & $2.556$ & \multicolumn{2}{c}{\nodata} & \multicolumn{2}{c}{\nodata} & \multicolumn{2}{c}{\nodata} \\
GRB 970828 & $-4.248$ & $94.936$ & $0.864$ & $0.112$ & $66.208$ & $2.781$ & $7.680$ & $0.161$ & $17.792$ & $0.310$ & \multicolumn{2}{c}{\nodata} & \multicolumn{2}{c}{\nodata} & \multicolumn{2}{c}{\nodata} \\
GRB 971214 & $-9.060$ & $16.564$ & $-9.060$ & $2.082$ & $15.892$ & $2.133$ & $-3.172$ & $2.951$ & $6.724$ & $2.970$ & \multicolumn{2}{c}{\nodata} & \multicolumn{2}{c}{\nodata} & \multicolumn{2}{c}{\nodata} \\
GRB 990123 & $-17.312$ & $111.200$ & $1.600$ & $0.161$ & $62.016$ & $1.179$ & $7.904$ & $0.072$ & $26.336$ & $0.757$ & $0.681$ & $0.091$ & $0.619$ & $0.099$ & $0.165$ & $0.050$ \\
GRB 990506 & $-0.390$ & $164.742$ & $1.952$ & $0.041$ & $128.608$ & $0.654$ & $12.032$ & $0.088$ & $83.392$ & $2.565$ & \multicolumn{2}{c}{\nodata} & \multicolumn{2}{c}{\nodata} & \multicolumn{2}{c}{\nodata} \\
GRB 990510 & $-0.320$ & $69.568$ & $0.688$ & $0.186$ & $55.888$ & $8.108$ & $38.976$ & $1.735$ & $5.760$ & $1.745$ & \multicolumn{2}{c}{\nodata} & \multicolumn{2}{c}{\nodata} & \multicolumn{2}{c}{\nodata} \\
GRB 990705 & $-1.698$ & $67.746$ & $1.648$ & $0.066$ & $33.232$ & $1.120$ & $7.488$ & $0.096$ & $14.720$ & $0.211$ & $0.053$ & $0.016$ & $0.103$ & $0.063$ & $0.016$ & $0.014$ \\
GRB 990712 & $-1.637$ & $18.821$ & $-1.637$ & $0.862$ & $16.629$ & $1.777$ & $0.784$ & $0.173$ & $10.784$ & $0.470$ & \multicolumn{2}{c}{\nodata} & \multicolumn{2}{c}{\nodata} & \multicolumn{2}{c}{\nodata} \\
\enddata
\tablecomments{A positive value of the spectral lag $\tau_\textrm{lag}$ corresponds to a delay of the soft photons.\\This table is available in its entirety in a machine-readable form in the online journal. A portion is shown here for guidance regarding its form and content.}

\end{deluxetable}

\clearpage
\begin{deluxetable}{lccrcrcrrc}
\rotate
\tabletypesize{\scriptsize}
\tablewidth{0pt}
\tablecaption{Spectral parameters \label{spectraltab}}
\tablehead{
\colhead{Burst}
& \colhead{Spec.}
& \colhead{$t_\textrm{start}$}
& \colhead{$\Delta T$}
& \colhead{Model}
& \colhead{$\alpha$}
& \colhead{$\beta$}
& \colhead{$E_\textrm{p}$}
& \colhead{$F$}
& \colhead{$\chi ^{2}/\textrm{d.o.f.}$}
\\
\colhead{Name}
& \colhead{type}
& \colhead{(s)}
& \colhead{(s)}
& \colhead{}
& \colhead{}
& \colhead{}
& \colhead{(keV)}
& \colhead{($10^{-6}$~erg~cm$^{-2}$~s$^{-1}$)}
& \colhead{(Prob.)}
}
\startdata
GRB 970228 & i & 0.000 & 8.448 & CPL\tablenotemark{\dag} & $-1.27_{-0.22}^{+0.24}$ & \nodata & $165_{-25}^{+39}$ & $0.53_{-0.05}^{+0.06}$ & 44.9/56 (0.86)\\
 & i &  &  & Band & $-1.24_{-0.22}^{+0.31}$ & $-2.82_{-7.18}^{+0.60}$ & $159_{-36}^{+38}$ & $0.60_{-0.06}^{+0.06}$ & 44.5/55 (0.84)\\
 & p & 0.000 & 0.256 & CPL\tablenotemark{\dag} & $-0.81_{-0.27}^{+0.32}$ & \nodata & $309_{-60}^{+102}$ & $3.38_{-0.40}^{+0.42}$ & 13.3/24 (0.96)\\
 & p &  &  & Band & $-0.76_{-0.29}^{+0.49}$ & $-2.64_{-7.36}^{+0.70}$ & $286_{-105}^{+102}$ & $4.14_{-0.47}^{+0.47}$ & 13.0/23 (0.95)\\
GRB 970828 & i & 0.000 & 70.656 & CPL & $-0.86_{-0.04}^{+0.04}$ & \nodata & $346_{-17}^{+19}$ & $0.89_{-0.03}^{+0.03}$ & 72.6/66 (0.27)\\
 & i &  &  & Band\tablenotemark{\dag} & $-0.73_{-0.06}^{+0.06}$ & $-2.18_{-0.15}^{+0.11}$ & $271_{-22}^{+24}$ & $1.33_{-0.05}^{+0.05}$ & 63.5/65 (0.53)\\
 & p & 17.920 & 5.120 & CPL & $-0.91_{-0.05}^{+0.05}$ & \nodata & $355_{-25}^{+29}$ & $2.20_{-0.10}^{+0.10}$ & 68.6/79 (0.79)\\
 & p &  &  & Band\tablenotemark{\dag} & $-0.78_{-0.07}^{+0.08}$ & $-2.18_{-0.15}^{+0.12}$ & $271_{-28}^{+31}$ & $3.23_{-0.26}^{+0.27}$ & 57.6/78 (0.96)\\
\enddata
\tablenotetext{\dag} {indicates the BEST model for the spectrum.}
\tablecomments{This table is available in its entirety in a machine-readable form in the online journal. A portion is shown here for guidance regarding its form and content.}

\end{deluxetable}

\setcounter{cit}{0}
\renewcommand{\cititem}[1]{\refstepcounter{cit}(\arabic{cit})~\label{En:#1}\citealt{#1}}

\begin{deluxetable}{llrrrrrrrrrcc}
\rotate
\tabletypesize{\scriptsize}
\tablewidth{0pt}
\tablecaption{Burst energetics\label{energytab}}
\tablehead {
\colhead{Burst name}
& \colhead{$z$}
& \colhead{$S$\tablenotemark{a}}
& \colhead{$T_{\rm peak,1024}$\tablenotemark{b}}
& \colhead{$F_{\rm peak,1024}$\tablenotemark{c}}
& \colhead{$T_{\rm peak,64}$\tablenotemark{b}}
& \colhead{$F_{\rm peak,64}$\tablenotemark{c}}
& \colhead{$T_{\rm peak,64,r}$\tablenotemark{b}}
& \colhead{$F_{\rm peak,64,r}$\tablenotemark{c}}
& \colhead{$E_{\rm iso}$\tablenotemark{d}}
& \colhead{$L_{\rm iso}$\tablenotemark{e}}
& \colhead{$F_{\rm lim}$\tablenotemark{c}}
& \colhead{$z_{\rm max}$}
}
\startdata
GRB 970228 & 0.69 & $8.07^{+0.49}_{-0.41}$ & -0.512\tablenotemark{\dag} & $2.24^{+0.43}_{-0.34}$ & 0.106 & $4.59^{+0.80}_{-0.78}$ & 0.118 & $4.12^{+0.66}_{-0.64}$ & $12.01 \pm 0.93$ & $9.6 \pm 1.5$ & 0.92 & 1.32 \\
GRB 970828 & 0.96 & $101.8^{+3.5}_{-3.6}$ & 20.256 & $4.30^{+0.46}_{-0.44}$ & 20.416 & $6.16^{+0.88}_{-0.87}$ & 20.416 & $6.11^{+0.71}_{-0.71}$ & $262.2 \pm 9.1$ & $30.9 \pm 3.6$ & 1.1 & 1.85 \\
GRB 971214 & 3.42 & $5.72^{+0.24}_{-0.22}$ & 2.752 & $0.604^{+0.071}_{-0.068}$ & 5.664 & $0.99^{+0.25}_{-0.25}$ & -0.280 & $0.69^{+0.12}_{-0.11}$ & $146.3 \pm 6.1$ & $78 \pm 13$ & 0.44 & 4.05 \\
GRB 990123 & 1.60 & $310.6^{+8.0}_{-7.8}$ & 5.872 & $25.6^{+1.1}_{-1.1}$ & 6.048 & $29.1^{+2.3}_{-2.3}$ & 5.984 & $27.7^{+1.6}_{-1.6}$ & $2133 \pm 54$ & $490 \pm 29$ & 3.1 & 5.04 \\
GRB 990506 & 1.31 & $261.0^{+8.8}_{-8.8}$ & 87.040 & $9.32^{+0.47}_{-0.46}$ & 90.048 & $13.8^{+1.1}_{-1.1}$ & 87.360 & $12.13^{+0.69}_{-0.68}$ & $1255 \pm 43$ & $134.2 \pm 7.6$ & 0.92 & 3.80 \\
GRB 990510 & 1.62 & $21.72^{+0.89}_{-0.82}$ & 44.160 & $3.56^{+0.35}_{-0.33}$ & 44.224 & $5.57^{+0.80}_{-0.78}$ & 44.544 & $4.35^{+0.52}_{-0.50}$ & $174.2 \pm 8.1$ & $81.0 \pm 9.6$ & 0.77 & 3.37 \\
GRB 990705 & 0.84 & $109.0^{+3.8}_{-3.8}$ & 14.000 & $4.80^{+0.44}_{-0.42}$ & 15.856 & $8.92^{+0.96}_{-0.96}$ & 15.824 & $8.36^{+0.76}_{-0.76}$ & $218.1 \pm 7.7$ & $30.7 \pm 2.8$ & 0.84 & 2.02 \\
GRB 990712 & 0.43 & $6.25^{+0.38}_{-0.33}$ & 10.880 & $0.802^{+0.118}_{-0.087}$ & 11.456 & $1.10^{+0.17}_{-0.17}$ & 11.504 & $1.05^{+0.14}_{-0.14}$ & $3.86 \pm 0.28$ & $1.20 \pm 0.18$ & 1.1 & 0.50 \\
\enddata
\tablenotetext{\dag}{Since the trigger mode light curve begins at $T_0 - 0.512$~s, the corresponding peak energy flux may be underestimated due to the absence of high-res data before $T_0 - 0.512$~s.}
\tablecomments{This table is available in its entirety in a machine-readable form in the online journal. A portion is shown here for guidance regarding its form and content.}

\tablenotetext{a} {In units of $10^{-6}$~erg~cm$^{-2}$.}
\tablenotetext{b}{The start time of the time interval when the peak count rate is reached, s.}
\tablenotetext{c} {In units of $10^{-6}$~erg~cm$^{-2}$~s$^{-1}$}.
\tablenotetext{d} {In units of $10^{51}$~erg.}
\tablenotetext{e} {In units of $10^{51}$~erg~s$^{-1}$.}
\end{deluxetable}

\begin{deluxetable}{llcc r@{$\pm$}l r@{$\pm$}l r@{$\pm$}l r@{$\pm$}l}
\tabletypesize{\scriptsize}
\tablewidth{0pt}
\tablecaption{Collimation-corrected parameters\label{collimationtab}}
\tablehead {
\colhead{Burst}
& \colhead{$t_\textrm{jet}$}
& \colhead{CBM}
& \colhead{Ref.\tablenotemark{a}}
& \multicolumn{2}{c}{$\theta_\textrm{jet}$}	
& \multicolumn{2}{c}{Collimation}
& \multicolumn{2}{c}{$E_{\gamma}$}
& \multicolumn{2}{c}{$L_{\gamma}$}
\\
\colhead{name}
& \colhead{(days)}
& \colhead{}
& \colhead{}
& \multicolumn{2}{c}{(deg)}
& \multicolumn{2}{c}{factor ($\times 10^{-3}$)}
& \multicolumn{2}{c}{($10^{49}$~erg)}
& \multicolumn{2}{c}{($10^{49}$~erg~s$^{-1}$)}
}

\startdata
GRB~990123 & $2.06_{-0.83}^{+0.83}$ & WM & (\ref{En:Zeh2006}) & 1.91 & 0.19 & 0.55 & 0.12 & 118.10 & 24.10 & 27.16 & 5.74 \\
GRB~990510 & $1.31_{-0.07}^{+0.07}$ & HM & (\ref{En:Zeh2006}) & 4.21 & 0.09 & 2.70 & 0.11 & 47.07 & 3.20 & 21.90 & 0.95 \\
GRB~990705 & 1\tablenotemark{b} & HM\tablenotemark{c} & (\ref{En:Masetti2000}) & 4.23 & 0.32 & 2.72 & 0.42 & 59.31 & 9.13 & 8.36 & 1.26 \\
 & & WM &  & 3.07 & 0.16 & 1.43 & 0.15 & 31.28 & 3.21 & 4.41 & 0.41 \\
GRB~990712 & $1.61_{-0.19}^{+0.19}$ & HM\tablenotemark{c} & (\ref{En:Bjornsson2001}) & 9.20 & 0.42 & 12.90 & 1.19 & 4.96 & 0.68 & 1.54 & 0.28 \\
 & & WM &  & 10.10 & 0.35 & 15.50 & 1.09 & 5.98 & 0.60 & 1.85 & 0.34 \\
GRB~991216 & $1.1_{-0.13}^{+0.13}$ & WM & (\ref{En:Zeh2006}) & 2.16 & 0.06 & 0.71 & 0.04 & 63.13 & 3.88 & 36.32 & 2.68 \\
GRB~000301C & $7.3_{-0.5}^{+0.5}$ & HM & (\ref{En:Berger2000}) & 9.33 & 0.30 & 13.20 & 0.85 & 44.56 & 7.20 & 74.95 & 8.08 \\
\enddata
\tablenotetext{a}{In cases where two references are given, the first one corresponds to the $t_\textrm{jet}$ estimate and the second one corresponds to the preferred CBM.}
\tablenotetext{b}{When no $t_\textrm{jet}$ uncertainty is available from the literature, we take the sample-mean $\sim 0.2 \cdot t_\textrm{jet}$ as a 68\% $t_\textrm{jet}$ error for the calculations.}
\tablenotetext{c}{In cases where no preferred CBM density profile is available from the literature, we provide the estimates for both HM and WM.   
}
\tablecomments{This table is available in a machine-readable form in the online journal.}
\tablerefs{
\cititem{Zeh2006}; \cititem{Masetti2000}; \cititem{Bjornsson2001}; \cititem{Berger2000}; \cititem{Galama2003}; \cititem{Price2003a}; \cititem{Resmi2005}; \cititem{Ghirlanda2007}; \cititem{Blustin2006}; \cititem{Cenko2006a}; \cititem{Soderberg2006}; \cititem{Schulze2011}; \cititem{Page2007}; \cititem{Chandra2008}; \cititem{Kann2007}; \cititem{Tanvir2010}; \cititem{Cenko2011}; \cititem{Cano2011}; \cititem{Filgas2011}; \cititem{Kann2011}; \cititem{Maselli2014}; \cititem{Fong2014}; \cititem{Nappo2016}; 
}

\end{deluxetable}

\begin{deluxetable}{lccccccc}
\tabletypesize{\scriptsize}
\tablewidth{0pt}
\tablecaption{Parameter statistics\label{stattab}}
\tablehead {
\colhead{Parameter}
& \colhead{Min}
& \colhead{Max}
& \colhead{Mean}
& \colhead{Median}
\\
\colhead{Name}
& \colhead{Value}
& \colhead{Value}
& \colhead{Value}
& \colhead{Value}
}
\startdata
Redshift & 0.096 & 5 & 1.50 & 1.32\\
$T_{100}$~(s) & 0.124 & 484.858 & 67.689 & 37.312\\
$T_{90}$~(s) & 0.070 & 440.826 & 46.557 & 21.664\\
$T_{50}$~(s) & 0.034 & 167.290 & 16.959 & 7.616\\
$T_\textrm{100,z}$~(s) & 0.088 & 170.884 & 29.476 & 13.974\\
$T_\textrm{90,z}$~(s) & 0.052 & 121.954 & 19.447 & 9.677\\
$T_\textrm{50,z}$~(s) & 0.025 & 49.733 & 7.220 & 3.002\\
$\tau_\textrm{lagG2G1}$~(ms) & 0.6 & 2495 & 292 & 150\\
\enddata

\end{deluxetable}

\begin{deluxetable}{lcccccc}
\tabletypesize{\scriptsize}
\tablewidth{0pt}
\tablecaption{LF and EF fits with BPL and Cutoff PL\label{tabLFEF}}
\tablehead {
\colhead{Data}
& \colhead{Evolution}
& \colhead{Model}
& \colhead{$\chi ^{2}(\textrm{d.o.f.})$}
& \colhead{$\alpha_1$}
& \colhead{$\alpha_2$}
& \colhead{$\textrm{log}$~$x_b$}
\\
\colhead{}
& \colhead{(PL index)}
& \colhead{}
& \colhead{}
& \colhead{}
& \colhead{}
& \colhead{($\textrm{log}$~$x_\textrm{cut}$)}
}
\startdata
$\psi(L')            $ & $\delta_L$=1.7 & BPL & 2.05 (133)& $-0.47 \pm 0.06$ & $-1.05 \pm 0.11$ & $0.27 \pm 0.12$\\
$\psi(L')            $ & $\delta_L$=1.7 & CPL & 18.5 (134)& $-0.60 \pm 0.04$ &                  & $2.10 \pm 0.15$\\
$\psi(E')            $ & $\delta_E$=1.1 & BPL & 19.2 (126)& $-0.36 \pm 0.01$ & $-1.28 \pm 0.11$ & $1.30 \pm 0.04$\\
$\psi(E')            $ & $\delta_E$=1.1 & CPL & 12.7 (127)& $-0.31 \pm 0.02$ &                  & $2.09 \pm 0.04$\\
\hline
$\psi(L_\textrm{iso})$ & no evolution   & BPL & 2.32 (133)& $-0.48 \pm 0.06$ & $-1.00 \pm 0.10$ & $0.96 \pm 0.15$\\
$\psi(L_\textrm{iso})$ & no evolution   & CPL & 8.90 (134)& $-0.54 \pm 0.04$ &                  & $2.58 \pm 0.11$\\
$\psi(E_\textrm{iso})$ & no evolution   & BPL & 17.2 (126)& $-0.35 \pm 0.01$ & $-1.29 \pm 0.12$ & $1.80 \pm 0.05$\\
$\psi(E_\textrm{iso})$ & no evolution   & CPL & 15.4 (127)& $-0.32 \pm 0.01$ &                  & $2.63 \pm 0.04$\\
\enddata
\end{deluxetable}

\begin{deluxetable}{lcccccccc}
\tabletypesize{\scriptsize}
\tablewidth{0pt}
\tablecaption{Hardness-intensity correlations\label{correlationtab}}
\tablehead {
\colhead{Correlation}
& \colhead{$N$}
& \colhead{$\rho_S$}
& \colhead{$P_{\rho_S}$}
& \colhead{$a$}
& \colhead{$b$}
& \colhead{$a_{\sigma_\textrm{int}}$}
& \colhead{$b_{\sigma_\textrm{int}}$}
& \colhead{$\sigma_\textrm{int}$}
}
\startdata
\cutinhead{Type~I GRBs}
$E_\textrm{p,i}$ vs $S$ & $12$ & $0.74$ & $5.8 \times 10^{-3}$ & $0.408 \pm 0.043$ & $4.98 \pm 0.22$ & $0.496 \pm 0.117$ & $5.52 \pm 0.62$ & $0.135$\\
$E_\textrm{p,i,$z$}$ vs $E_{\textrm{iso}}$ & $12$ & $0.83$ & $9.5 \times 10^{-4}$ & $0.364 \pm 0.030$ & $-15.70 \pm 1.53$ & $0.266 \pm 0.068$ & $-10.61 \pm 3.47$ & $0.181$\\
$E_\textrm{p,p}$ vs $F_\textrm{peak}$ & $12$ & $0.54$ & $7.1 \times 10^{-2}$ & $0.340 \pm 0.045$ & $4.39 \pm 0.19$ & $0.349 \pm 0.161$ & $4.52 \pm 0.74$ & $0.188$\\
$E_\textrm{p,p,$z$}$ vs $L_\textrm{iso}$   & $12$ & $0.67$ & $1.7 \times 10^{-2}$ & $0.396 \pm 0.034$ & $-17.68 \pm 1.78$ & $0.243 \pm 0.078$ & $-9.61 \pm 4.07$ & $0.200$\\

\cutinhead{Type~II GRBs}
$E_\textrm{p,i}$ vs $S$ & $137$ & $0.59$ & $3.7 \times 10^{-14}$ & $0.418 \pm 0.002$ & $4.06 \pm 0.01$ & $0.295 \pm 0.031$ & $3.66 \pm 0.14$ & $0.227$\\
$E_\textrm{p,i,$z$}$ vs $E_{\textrm{iso}}$ & $137$ & $0.70$ & $1.4 \times 10^{-21}$ & $0.469 \pm 0.003$ & $-22.35 \pm 0.14$ & $0.338 \pm 0.026$ & $-15.27 \pm 1.37$ & $0.229$\\
$E_\textrm{p,p}$ vs $F_\textrm{peak}$ & $136$ & $0.58$ & $2.2 \times 10^{-13}$ & $0.453 \pm 0.004$ & $4.68 \pm 0.02$ & $0.363 \pm 0.041$ & $4.31 \pm 0.21$ & $0.253$\\
$E_\textrm{p,p,$z$}$ vs $L_\textrm{iso}$   & $136$ & $0.73$ & $1.6 \times 10^{-23}$ & $0.494 \pm 0.005$ & $-23.32 \pm 0.26$ & $0.347 \pm 0.029$ & $-15.52 \pm 1.51$ & $0.251$\\
\enddata

\tablecomments{$N$ is the number of bursts in the fit sample, $\rho_S$ is the Spearman correlation coefficient, $P_{\rho_S}$ is the corresponding chance probability, $a$ ($a_{\sigma_\textrm{int}}$) and $b$ ($b_{\sigma_\textrm{int}}$) are the slope and the intercept for the fits without (with) intrinsic scatter $\sigma_\textrm{int}$.}
\end{deluxetable}

\end{document}